\documentclass[prd,superscriptaddress,nofootinbib,amsmath,amssymb,aps,11pt]{revtex4-2}

\usepackage{bm}
\usepackage{amsfonts}
\usepackage{latexsym}
\usepackage[latin1]{inputenc}
\usepackage{graphicx}
\usepackage{amsmath}
\usepackage{palatino}
\usepackage{mathpazo}
\usepackage[normalem]{ulem}

\usepackage{booktabs}
\usepackage{dcolumn}
\usepackage{natbib}
\usepackage[dvipsnames]{xcolor}

\begin{document}
	\title{$3+1$ non-linear evolution of Ricci-coupled scalar-Gauss-Bonnet gravity}
	
	\author{Daniela D. Doneva}
	\email{daniela.doneva@uni-tuebingen.de}
	\affiliation{Theoretical Astrophysics, Eberhard Karls University of T\"ubingen, T\"ubingen 72076, Germany}
	\affiliation{INRNE - Bulgarian Academy of Sciences, 1784  Sofia, Bulgaria}

	\author{Llibert Arest\'e Sal\'o}
	\email{l.arestesalo@qmul.ac.uk}
	\affiliation{School of Mathematical Sciences, Queen Mary University of London, Mile End Road, London, E1 4NS, United Kingdom}
	
	\author{Stoytcho S. Yazadjiev}
	\email{yazad@phys.uni-sofia.bg}
	\affiliation{Theoretical Astrophysics, Eberhard Karls University of T\"ubingen, T\"ubingen 72076, Germany}
	\affiliation{Department of Theoretical Physics, Faculty of Physics, Sofia University, Sofia 1164, Bulgaria}
	\affiliation{Institute of Mathematics and Informatics, 	Bulgarian Academy of Sciences, 	Acad. G. Bonchev St. 8, Sofia 1113, Bulgaria}
	
\begin{abstract}
	Scalar-Gauss-Bonnet (sGB) gravity with an additional coupling between the scalar field and the Ricci scalar exhibits very interesting properties related to black hole stability, evasion of binary pulsar constraints, and general relativity as a late-time cosmology attractor. Furthermore, it was demonstrated that a spherically symmetric collapse is well-posed for a wide range of parameters. In the present paper we examine further the well-posedness through $3+1$ evolution of static and rotating black holes. We show that the evolution is indeed hyperbolic if the weak coupling condition is not severely violated. The loss of hyperbolicity is caused by the gravitational sector of the physical modes, thus it is not an artifact of  the gauge choice. We further seek to compare the Ricci-coupled sGB theory against the standard sGB gravity with additional terms in the Gauss-Bonnet coupling. We find strong similarities in terms of well-posedness, but we also point out important differences in the stationary solutions. As a byproduct, we show strong indications that stationary near-extremal scalarized black holes exist within the Ricci-coupled sGB theory, where the scalar field is sourced by the spacetime curvature rather than the black hole spin.
\end{abstract}

\maketitle

 \section{Introduction}
	The rapid advance of gravitational wave detectors gives us confidence that in the next decades we will have the necessary observations to allow for precise tests of gravity \cite{LIGOScientific:2016lio,LIGOScientific:2020tif,LISA:2022kgy,Barack:2018yly}. The availability of accurate theoretical gravitational waveforms both in general relativity (GR) and its modifications will be crucial for the correct interpretation of the detected gravitational wave events. While the accuracy of the former is on a steady path of improvement, numerical relativity simulations beyond GR are still in the development phase. The reason behind that is twofold -- first, the field equations typically get increasingly more complicated compared to GR when we start modifying the original GR action. The second more fundamental obstacle is the question of well-posedness, which is one of the main building blocks if we want to be able to perform time evolution. Even though strong hyperbolicity is proven in certain formulations of the $3+1$ field equations in GR \cite{Sarbach:2002bt,Beyer:2004sv,Reula:2004xd}, it is by no means guaranteed that this will remain true in modified gravity \cite{Papallo:2017qvl,Bernard:2019fjb}. Whether this is sourced by an intrinsic problem of the theory or just a gauge change is required, is a question that awaits answer in a number of GR modifications.
	
	Our focus in the present paper will be the scalar-Gauss-Bonnet gravity (sGB). It provides an important playground for studying the possible deviations from GR one can have in an effective field theory of gravity while keeping the field equations of second order. It gained particular attention because it was proven that black holes with scalar hair can exist within this theory \cite{Mignemi:1992nt,Kanti:1995vq,Torii:1996yi,Pani:2009wy,Sotiriou:2013qea}, including spontaneously scalarized ones \cite{Doneva:2017bvd,Silva:2017uqg,Antoniou:2017acq}. Even though evolution in spherical symmetry can be hyperbolic for a weak enough coupling \cite{Ripley:2019hxt,R:2022hlf}, solving the full $3+1$ field equations is much more subtle with the standard harmonic gauge proven to be non-well-posed \cite{Papallo:2017qvl}. Interestingly, loss of hyperbolicity is observed even at the level of linear perturbations \cite{Blazquez-Salcedo:2018jnn,Blazquez-Salcedo:2020rhf}.  
	
	An important breakthrough was the proof that a modified harmonic gauge leads to well-posed $3+1$ field equations in the weak coupling regime, i.e. when the contributions of the Gauss-Bonnet term to the field equations are smaller than the two-derivative Einstein-scalar field terms  
 \cite{Kovacs:2020pns,Kovacs:2020ywu}. This regime is exactly where sGB gravity can be considered as a viable effective field theory. This eventually allowed the development of $3+1$ numerical relativity codes \cite{East:2021bqk,Corman:2022xqg} and an extension to a well-posed modified puncture gauge \cite{AresteSalo:2022hua,AresteSalo:2023mmd}.
	
	The search for a well-posed formulation of sGB gravity also eventually ignited the interest in alternative ways to address the problem. An interesting approach is to ``fix'' the field equations, which can be regarded as providing a weak completion of the considered effective field theory \cite{Cayuso:2017iqc,Franchini:2022ukz,Cayuso:2023xbc,Lara:2024rwa} and is inspired by the dissipative relativistic hydrodynamics \cite{Muller:1967zza,Israel:1976efz,Israel:1976tn}. Even though it seems promising, further development is needed to have a self-consistent and robust $3+1$ evolution. Another approach is to modify the original sGB action and a natural extension is to add a coupling between the scalar field and the Ricci scalar coupling \cite{Antoniou:2021zoy}. It was shown that it can lead to hyperbolic evolution in the case of a spherically symmetric collapse of a scalar field \cite{Thaalba:2023fmq}. As a matter of fact, this theory has other interesting features such as linear stability of black holes that are otherwise unstable in the standard sGB gravity and the possibility to evade binary pulsar constraints for certain ranges of parameters \cite{Ventagli:2021ubn}. Furthermore, it also cures some of the problems encountered when treating scalar-tensor theories in a cosmological set-up. In particular, adding the Ricci scalar coupling is a minimal model which succeeds in having GR as a cosmological late-time attractor \cite{Antoniou:2020nax}, which is otherwise not true \cite{Anderson:2016aoi,Damour:1992kf,Franchini:2019npi}. However, other problems occurring at early times persist within this model \cite{Anson:2019uto} and one would need to add extra operators \cite{Babichev:2024txe} in order to cure them.  
	
 In the present paper, we aim to explore further the Ricci-coupled sGB gravity by investigating the well-posedness of the theory in the $3+1$ formulation of the field equations using the modified gauge proposed in \cite{Kovacs:2020pns,Kovacs:2020ywu} in the puncture gauge approach \cite{AresteSalo:2022hua,AresteSalo:2023mmd}. We also compare the theory with certain subclasses of sGB gravity known to also lead to linearly stable black holes \cite{Minamitsuji:2018xde,Silva:2018qhn} and hyperbolic $3+1$ evolution (for weak enough scalar fields) \cite{Ripley:2019hxt,Doneva:2023oww}. 
	
	The paper is organized as follows. In Section \ref{sec:theory} we define the Ricci-coupled scalar-Gauss-Bonnet theory considered in this work and give a brief overview of the modified CCZ4 formalism, together with the concepts of the effective metric and the weak coupling conditions, which are relevant in this manuscript. In Section \ref{sec:results}, after motivating the specific form of the coupling functions that we are using and introducing the numerical set-up, we present our results regarding the non-linear evolution of rotating and non-rotating black holes and their comparison within two different set of coupling functions. Finally, in Appendix \ref{app:eom} we include the equations of motion of the theory in the modified CCZ4 formalism and in Appendix \ref{app:convergence} we test the validity of the developed code.

 We follow the conventions in Wald's book \cite{Wald:1984rg}. Greek letters $\mu, \nu, ...$ denote spacetime
indices and they run from $0$ to $3$; Latin
letters $i, j, ...$ denote indices on the spatial hypersurfaces and they run from $1$ to $3$. We
set $G=c=1$.
	
	\section{Theoretical background}\label{sec:theory}
	
	We consider a scalar-Gauss-Bonnet theory with a Ricci coupling (which belongs to the Horndeski class) corresponding to the following action,
	\begin{equation}\label{eq:action}
		S=\frac{1}{16\pi}\int d^4x\sqrt{-g}\Big(R+X-\beta(\varphi)R+\frac{\lambda(\varphi)}{4}{\mathcal R}^2_{\text{GB}}\Big)\,,
	\end{equation}
	where $R$ is the Ricci scalar with respect to the spacetime metric $g^{\mu\nu}$, ${\mathcal R}^2_{\text{GB}}$ is the Gauss-Bonnet invariant defined as ${\mathcal R}^2_{\text{GB}}=R^2-4R_{\mu\nu}R^{\mu\nu}+R_{\mu\nu\rho\sigma}R^{\mu\nu\rho\sigma}$, $\varphi$ is the scalar field with $X=-\frac{1}{2}\nabla_{\mu}\varphi\nabla^{\mu}\varphi$ being its kinetic term. The Gauss-Bonnet and Ricci couplings are controlled by arbitrary functions of the scalar field with $\lambda(\varphi)$ having dimensions of $[\text{length}]^2$ and $\beta(\phi)$ being dimensionless. Its equations of motion yield
\begin{subequations}\label{FE}
     \begin{eqnarray}
		&&(1-\beta(\varphi))(R_{\mu\nu}- \frac{1}{2}R g_{\mu\nu}) + \Gamma_{\mu\nu}= \frac{1}{2}\nabla_\mu\varphi\nabla_\nu\varphi -  \frac{1}{4}g_{\mu\nu} \nabla_\alpha\varphi \nabla^\alpha\varphi +(g_{\mu\nu}\Box- \nabla_{\mu}\nabla_{\nu})\beta(\varphi)\,,\\
		&&\nabla_\alpha\nabla^\alpha\varphi=  -  \frac{\lambda'(\varphi)}{4} {\cal R}^2_{GB}+\beta'(\varphi)R\,,
	\end{eqnarray}
\end{subequations}	
	where   $\Gamma_{\mu\nu}$ is defined as 	
	\begin{eqnarray}
		\Gamma_{\mu\nu}&=& -\frac{1}{2}R\Omega_{\mu\nu} - \Omega_{\alpha}^{~\alpha}\left(R_{\mu\nu} - \frac{1}{2}R g_{\mu\nu}\right) 
		+ 2\,R_{\alpha(\mu}\Omega^{~\alpha}_{\nu)} - g_{\mu\nu} R^{\alpha\beta}\Omega_{\alpha\beta} 
		+ \,  R^{\beta}_{\;\mu\alpha\nu}\Omega^{~\alpha}_{\beta}\,,
	\end{eqnarray}  
	with $\Omega_{\mu\nu}= \nabla_{\mu}\nabla_{\nu}\,\lambda(\varphi)$.
	
	\subsection{Modified CCZ4 formalism}
 The equations of motion that follow from varying \eqref{eq:action} in the modified harmonic gauge introduced by
\cite{Kovacs:2020ywu,Kovacs:2020pns} and supplemented by constraint damping terms are given by \eqref{FE} with the following replacement,
    \begin{eqnarray}\label{mod_FE}
        R^{\mu\nu}-\tfrac{1}{2}R g^{\mu\nu}\to R^{\mu\nu}-\tfrac{1}{2}R g^{\mu\nu}+2\big(\delta_{\alpha}^{(\mu}\hat{g}^{\nu)\beta}-\tfrac{1}{2}\delta_{\alpha}^{\beta}\hat{g}^{\mu\nu}\big)\nabla_{\beta}Z^{\alpha}-\kappa_1\big[2n^{(\mu}Z^{\nu)}+\kappa_2n^{\alpha}Z_{\alpha}g^{\mu\nu} \big]\,,
    \end{eqnarray}
    where $\hat{g}^{\mu\nu}$ and $\tilde{g}^{\mu\nu}$ are two auxiliary Lorentzian metrics that ensure that gauge modes and gauge condition violating modes propagate at distinct speeds from physical modes, as in \cite{Kovacs:2020pns, Kovacs:2020ywu} \footnote{Note that $\tilde{g}^{\mu\nu}$ is hidden in the definition of the constraints $Z^{\mu}$ (see \cite{AresteSalo:2022hua,AresteSalo:2023mmd} for further details).}. They can be defined as
    \begin{eqnarray}\label{eq:guage_choice}
        \tilde{g}^{\mu\nu}=g^{\mu\nu}-a(x)n^{\mu}n^{\nu}\,\qquad \hat{g}^{\mu\nu}=g^{\mu\nu}-b(x)n^{\mu}n^{\nu}\,,
    \end{eqnarray}
    where $a(x)$ and $b(x)$ are arbitrary functions such that $0<a(x)<b(x)$ and $n^{\mu}=\tfrac{1}{\alpha}(\delta_t^{\mu}-\beta^i\delta_i^{\mu})$ is the unit timelike vector normal to the $t\equiv x^0 = $const. hypersurfaces with $\alpha$ and $\beta^i$ being the lapse function and shift vector of the $3+1$ decomposition of the spacetime metric, namely
    \begin{eqnarray}
        ds^2=-\alpha^2dt^2+\gamma_{ij}(dx^i+\beta^idt)(dx^j+\beta^jdt)\,.
    \end{eqnarray}
   The damping terms in \eqref{mod_FE}, whose coefficients should satisfy $\kappa_1>0$ and $\kappa_2>-\tfrac{2}{2+b(x)}$, guarantee that constraint violating modes are exponentially suppressed \cite{AresteSalo:2022hua,AresteSalo:2023mmd}.

    In Appendix \ref{app:eom} we have written down the evolution equations for the $3+1$ formalism. The versions of the $1+log$ slicing and Gamma-driver evolution equations that result in the modified puncture gauge are
    \begin{eqnarray}
        \partial_t\alpha=\beta^i\partial_i\alpha-\tfrac{2\alpha}{1+a(x)}(K-2\Theta)\,,\\
        \partial_t\beta^i=\beta^j\partial_j\beta^i+\tfrac{3}{4}\tfrac{\hat{\Gamma}^i}{1+a(x)}-\tfrac{a(x)\alpha\partial_i\alpha}{1+a(x)}\,,
    \end{eqnarray}
    where $\Theta=Z^0$, $K$ is the trace of the extrinsic curvature of the induced metric $\gamma_{ij}$, $\hat{\Gamma}^i=\tilde{\gamma}^{kl}\tilde{\Gamma}^i_{kl}+2\tilde{\gamma}^{ij}Z_j$, with $\tilde{\Gamma}^i_{kl}$ being the Christoffel symbols associated to the conformal spatial metric $\tilde{\gamma}_{ij}\equiv\chi\gamma_{ij}$, where $\chi=\det(\gamma_{ij})^{-1/3}$.

	\subsection{Effective metric}

 The hyperbolicity of the equations of motion is held when its principal part is diagonalizable with real eigenvalues and a complete set of linearly independent and bounded eigenvectors that depend smoothly on the variables. The eigenvalues from the gauge sectors lie on the null cones of the auxiliary metrics, while the physical sector is described by a characteristic polynomial of degree 6 which factorises into a product of
quadratic and quartic polynomials \cite{Reall:2021voz}. The former is defined in terms of an ``effective metric'' and
is associated with a ``purely gravitational'' polarization, whereas the latter generically involves
a mixture of gravitational and scalar field polarizations. 

Even though the ``fastest'' degrees of freedom are associated with the quartic polynomial \cite{Reall:2021voz}, it is not necessarily the case
that hyperbolicity loss should occur first in their sector. Moreover, there is no simple way to study the hyperbolicity of that sector. This is why we have focused on the ``purely gravitational'' polarizations, which appear to coincide with the breakdown of the simulation as was seen in \cite{Doneva:2023oww} and in this work.

In the Ricci-coupled sGB theory, the effective metric yields
	\begin{equation} 
		g_{\text{eff}}^{\mu\nu}=g^{\mu\nu}(1-\beta(\varphi))-\Omega^{\mu\nu}\,,
	\end{equation}
	and its determinant (normalized to its value in pure GR) can be expressed as
	\begin{eqnarray}\label{eq:Geff_NonNormalized}
		\frac{\det(g^{\mu\nu}_{\text{eff}})}{\det(g^{\mu\nu})}&=&\left(\tfrac{1}{1+\Omega^{\perp\perp}-\beta(\varphi)}\right)^2\det\left\{\tfrac{1}{\chi}\big[(\gamma^{ij} (1-\beta(\varphi)) - \Omega^{ij})(1 + \Omega^{\perp\perp}- \beta(\varphi)) -\tfrac{2}{\alpha}\Omega^{\perp(i} \beta^{j)}\right.\nonumber
		\\&&\left. \hspace{3cm}- (1-\beta(\varphi))\Omega^{\perp\perp} \tfrac{\beta^i \beta^j}{\alpha^2}+ \Omega^{\perp i} \Omega^{\perp j}\big]\right\}\,,
	\end{eqnarray}
	where $\Omega^{ij}=\gamma^i_{\mu}\gamma^j_{\nu}\Omega^{\mu\nu}$, $\Omega^{\perp i}=-n_{\mu}\gamma^i_{\nu}\Omega^{\mu\nu}$ and $\Omega^{\perp\perp}=n_{\mu}n_{\nu}\Omega^{\mu\nu}$. In the results shown later we will consider the normalized determinant
 \begin{eqnarray}\label{eq:Geff}
     G_{\rm eff}\equiv (1+\Omega^{\perp\perp}-\beta(\varphi))^2\,\frac{\det(g^{\mu\nu}_{\text{eff}})}{\det(g^{\mu\nu})}\,,
 \end{eqnarray}
 which has no divergences when hyperbolicity is lost and is normalized to unity in the absence of any scalar field.
	
	\subsection{Weak coupling condition}

 One of the main reasons for the relevance of the sGB theory is that it accounts for the more general parity-invariant (up to field redefinitions)
scalar-tensor theory of gravity up to four derivatives, when considering a scalar field with no potential and neglecting the four-derivative scalar term, which
we see from our work in \cite{AresteSalo:2022hua,AresteSalo:2023mmd} is justified since it is always subdominant
to the effect of the Gauss-Bonnet term.

In this sense, we view these theories as effective field theories (EFTs) that arise as a
low energy limit of a more fundamental theory, whose terms are organised in a derivative expansion and appear multiplied by dimensionful coupling
constants that encode the effects of the underlying (unknown) microscopic theory.

Therefore, in order for our theory to be justified and valid as an EFT, one has to make sure that we are not beyond the threshold where the EFT breaks down and the higher derivative terms would become relevant. This is ensured as long as our theory is in the weak coupling regime throughout all its evolution. Namely, we require that the contributions of the Gauss-Bonnet term to the field equations are smaller than the two-derivative Einstein-scalar field terms, which can be expressed in the form of the following weak coupling condition \cite{AresteSalo:2023mmd,Doneva:2023oww}:
	\begin{eqnarray} \label{eq:wcc}
		\sqrt{|\lambda'(\varphi)|}/L \ll 1\,,
	\end{eqnarray}
	where $L$ accounts for any characteristic length
scale of the system associated to the spacetime curvature
and the gradients of the scalar field, which can be computed as
\begin{eqnarray}
    L^{-1}=\max\{|R_{ij}|^{1/2},|\nabla_{\mu}\varphi|,|\nabla_{\mu}\nabla_{\nu}\varphi|^{1/2},|{\mathcal R}^2_{\text{GB}}|^{1/4}\}\,.
\end{eqnarray}

	\section{Results} \label{sec:results}
	\subsection{Coupling functions}
	We will concentrate on the following forms of the coupling functions $\lambda(\varphi)$ and $\beta(\varphi)$:
	\begin{eqnarray}
		\lambda(\varphi) &=& \lambda_{\text{GB}}\, \varphi^2 + \gamma_{\text{GB}}\, \varphi^4, \label{eq:f_phi} \\
		\beta(\varphi) &=& \beta_{\text{Ricc}}\, \varphi^2.  \label{eq:beta_phi}      
	\end{eqnarray}
	As it is well known, a pure $\varphi^2$ term in the $\lambda(\varphi)$ function is enough to admit scalarized black holes \cite{Doneva:2022ewd,Doneva:2017bvd,Silva:2017uqg,Antoniou:2017acq}, but they are linearly unstable. The minimum modification to stabilize the solutions is to add a $\varphi^4$ term \cite{Minamitsuji:2018xde,Silva:2018qhn} or an alternative is to slightly change the theory, such as the introduction of a Ricci scalar coupling $\beta(\varphi)$ \cite{Antoniou:2021zoy} considered in the present paper. Thus, if $\gamma_{\text{GB}}$ and $\beta_{\text{Ricc}}$ are large enough by absolute value, the resulting black hole solutions are linearly stable.
	
	In the first part of the results presented below, we will consider the case of $\gamma_{\text{GB}}=0$ because our main goal is to examine the effect of the Ricci scalar coupling on the hyperbolicity. In the second part of the results, we will compare the effects of the $\gamma_{\text{GB}}\, \varphi^4$ term in $\lambda(\varphi)$ on the one hand and the Ricci coupling on the other. The motivation is that these are the simplest modifications of pure sGB gravity with a $\lambda(\varphi)=\lambda_{\text{GB}}\,\varphi^2$ coupling that lead to a restoration of stability and share similar properties of the solutions.
	
	\subsection{Numerical set-up and hyperbolicity loss treatment}
	It was shown in \cite{Thaalba:2023fmq} that the $1+1$ non-linear evolution within the Ricci-coupled sGB theory is hyperbolic for an extensive region of the parameter space. It is natural to generalize these results to a $3+1$ evolution where fixing a gauge is much more subtle. We have all reasons to believe that the modified gauge for sGB gravity \cite{Kovacs:2020pns,Kovacs:2020ywu} will also work for the considered theory with a Ricci scalar coupling. We also conjecture that, similarly to sGB gravity \cite{Reall:2021voz}, the loss of hyperbolicity, at least for the considered simulations, is related to the physical modes of the purely gravitational sector rather than the mixed scalar-gravitational one. This is based on the observation that the determinant of the effective metric \eqref{eq:Geff} turns negative right before the breakdown of the simulation. This is a signal that either the speed of these modes diverge or they become degenerate \cite{Reall:2021voz,Doneva:2023oww}.
	
	An important property of the sGB modified gauge proposed in \cite{Kovacs:2020pns,Kovacs:2020ywu} is that the mathematical proof for well-posedness is valid only in the weak coupling regime where the scalar field should be weak enough. As a matter of fact, in practice, hyperbolicity is also preserved when the weak coupling condition is slightly violated \cite{Doneva:2023oww}. It is natural to assume that this will also remain true when we consider an additional coupling between the Ricci scalar and the scalar field. This is why we have performed a series of numerical relativity simulations to probe the hyperbolicity of the employed Ricci-coupled sGB theory. A newly developed modification of \texttt{GRFolres} \cite{AresteSalo:2023hcp} (based on \texttt{GRChombo} \cite{Clough:2015sqa,Andrade:2021rbd,Radia:2021smk}) was implemented taking into account the Ricci scalar coupling in eqs. \eqref{FE}, which are explicitly written down in our modified CCZ4 formalism in Appendix \ref{app:eom}. Details about the code convergence and constraint violation are presented in  Appendix \ref{app:convergence}.
	
	We consider as initial data a Kerr black hole with a scalar field Gaussian pulse superimposed on it. In the theory we study, the Kerr black hole is a solution of the field equations but the parameters are chosen in such a way so that it is linearly unstable. As the evolution proceeds and the scalar pulse ``hits'' the black hole, the scalar hair starts developing  quickly until it reaches equilibrium or a loss of hyperbolicity occurs. The resolution that we mostly worked with is 128 points in each spatial direction with 6 refinement levels and a domain size of $256M$.
	
	The employed puncture gauge enables us to evolve the spacetime also through the black hole horizon, thus no explicit excision of the horizon is being made. Still, in order to achieve a stable evolution one has to ``turn-off'' the Gauss-Bonnet coupling inside the black hole \cite{AresteSalo:2022hua,AresteSalo:2023mmd} and practically evolve GR in the interior. As the apparent horizon is being approached, the Gauss-Bonnet term is gradually turned on so that outside the apparent horizon we are solving the full field equations. As long as the turning on and off of the sGB coupling is performed entirely inside the apparent horizon, this does not affect the spacetime outside the black hole. As a matter of fact, there is a second important reason for switching off the Gauss-Bonnet terms in the black hole interior. Typically, a hyperbolicity loss develops first inside the black hole horizon and then it can emerge above it \cite{Corelli:2022pio}. As long as this non-hyperbolic region is entirely inside the horizon, it is casually disconnected from the rest of the spacetime and it can be still accepted as a viable black hole solution. From the point of view of a numerical relativity code, though, as long as an elliptic region forms anywhere inside the computational domain it leads to unavoidable numerical divergences. Therefore, if we want to determine the threshold between hyperbolic and non-hyperbolic solutions (outside the apparent horizon) it is desirable to cure the black hole interior through the described procedure of switching on and off the Gauss-Bonnet terms.
	
	\subsection{Hyperbolicity of black hole non-linear evolution in Ricci-coupled sGB theory}
	\begin{figure}
		\centering
		\includegraphics[width=0.5\linewidth]{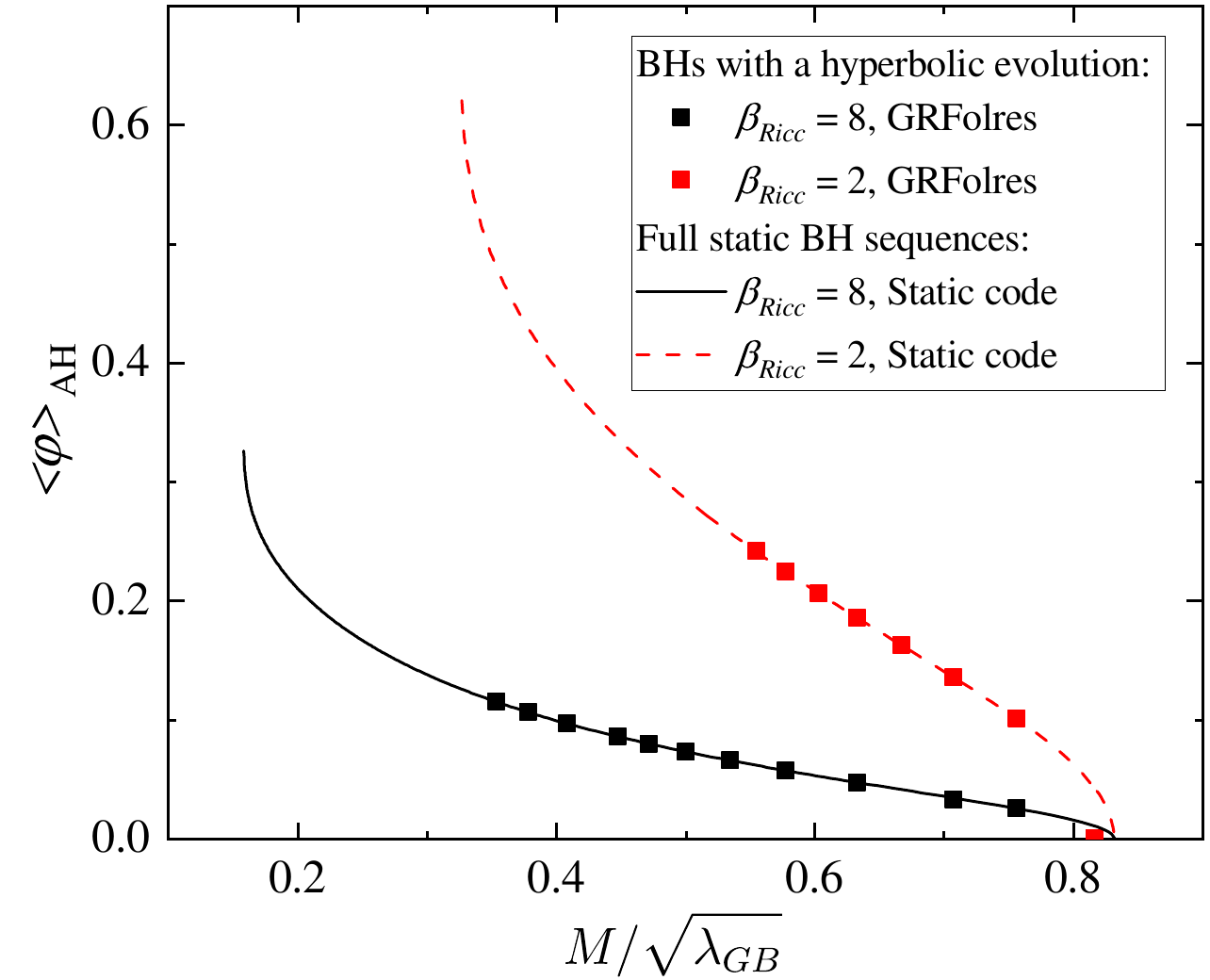}
		\caption{The mean value of the scalar field at the apparent horizon $\left<\varphi \right>_{\rm AH}$ as a function of the normalized black hole mass $M/\sqrt{\lambda_{\text{GB}}}$ for sequences of black holes with $\gamma_{\text{GB}}=0$ and two different values of the Ricci coupling constant $\beta_{\text{Ricc}}=2$ and $\beta_{\text{Ricc}}=8$. The squares are the end states of the $3+1$ simulations of black hole scalarization while the lines depict the full sequence of solutions obtained through solving the static field equations. The sequences of red and black squares are terminated at the last model for which we were able to perform hyperbolic evolution, since black hole evolutions with lower $M/\sqrt{\lambda_{\text{GB}}}$ develop hyperbolicity loss during the scalar field growth. }
		\label{fig:hyperbolic_seq}
	\end{figure}

 First, we start presenting the results for the evolution of sequences of non-rotating black holes with fixed $\gamma_{\text{GB}}=0$ and increasing mass. Two different values of the Ricci coupling constant $\beta_{\text{Ricc}}$ are considered, being adjusted in such a way that the resulting static black hole solutions are linearly stable. The mass and the scalar field at the black hole horizon are plotted in Fig. \ref{fig:hyperbolic_seq} for sequences of models at the end state of the numerical relativity simulations of black hole scalarization (after the metric and scalar field stabilize and become nearly static). Red and black squares in the figure correspond to the two values of $\beta_{\text{Ricc}}$. Naturally, only the models where the evolution is hyperbolic are depicted because typically hyperbolicity is lost at early times of the scalar field development \cite{Doneva:2023oww}. As evident in Fig. \ref{fig:hyperbolic_seq}, we could reach higher maximum scalar fields at the horizon (before hyperbolicity is lost) for the smaller value $\beta_{\text{Ricc}}=2$. On the other hand, the range of values of $M/\sqrt{\lambda_{\text{GB}}}$ where the black holes have a well-posed evolution enlarges with the increase of $\beta_{\text{Ricc}}$, which is consistent with the findings in spherical symmetry \cite{Thaalba:2023fmq}.

     As a comparison, with solid and dashed lines in Fig. \ref{fig:hyperbolic_seq} we plot the sequence of solutions resulting from solving the set of static field equations similar to \cite{Doneva:2017bvd}. Thus, the lines contain all asymptotically flat, regular, and linearly stable black hole solutions regardless of their hyperbolicity. The branches originate from $\left<\varphi \right>_{\rm AH}=0$, at the bifurcation point of the Schwarzschild solution, and they are terminated at some smaller $M/\sqrt{\lambda_{\text{GB}}}$. As one can see in the figure, the lines match very well the points, which is a strong argument for the correctness of the developed extension of \texttt{GRFolres}. As expected, the lines span a larger range of $M/\sqrt{\lambda_{\text{GB}}}$ as compared to the models resulting from non-linear evolution. The reason for that is that black holes with larger $\left<\varphi \right>_{\rm AH}$  cannot be formed dynamically through a hyperbolic time evolution, i.e. the dynamical variables diverge before the scalar field settles to a constant value. Therefore, similarly to pure sGB gravity \cite{Doneva:2023oww}, only the small scalar field black hole solutions are hyperbolic.

 	\begin{figure}
		\centering
		\includegraphics[width=1\linewidth]{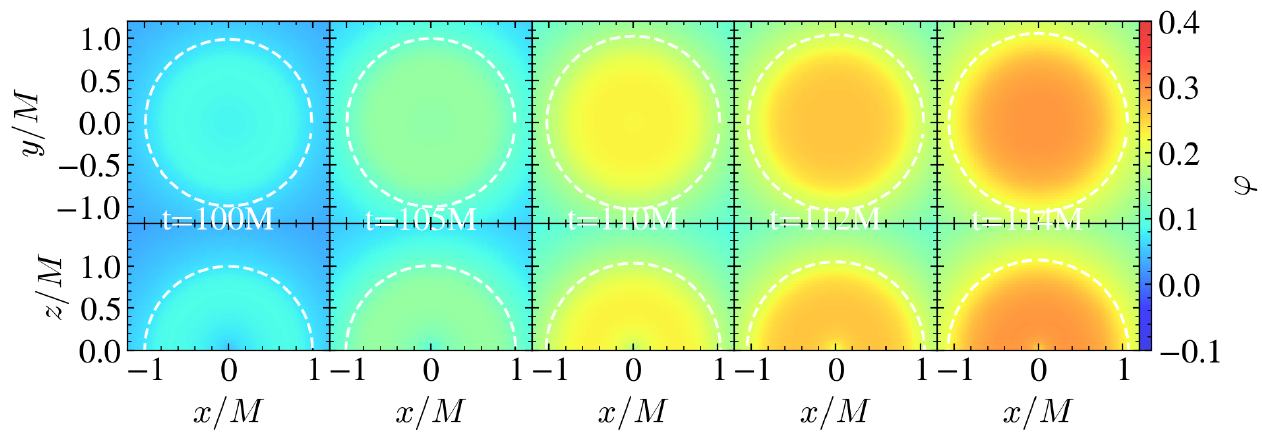}
		\includegraphics[width=1\linewidth]{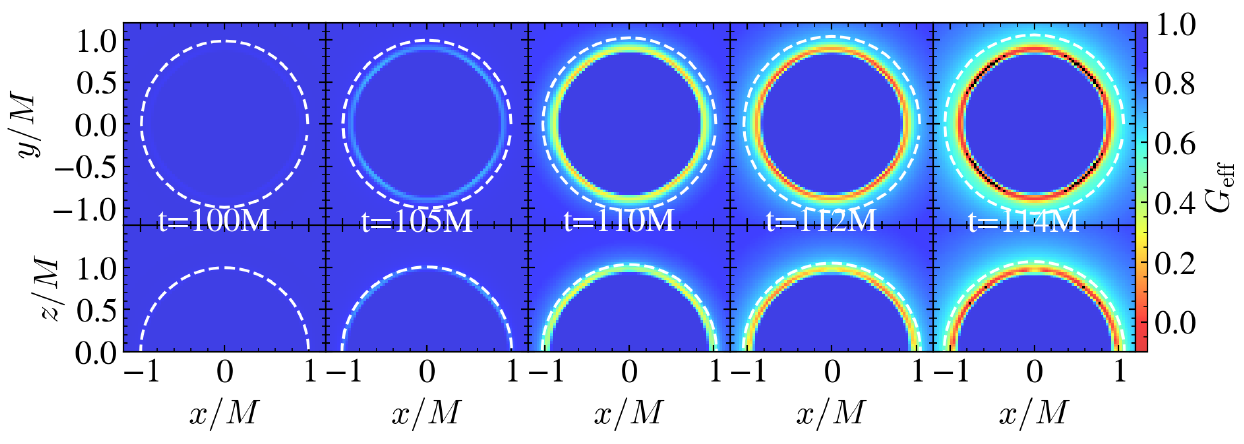}
		\caption{Time evolution of a non-rotating black hole with $\lambda_{\text{GB}}/M^2=4$, $\gamma_{\text{GB}}=0$, $\beta_{\text{Ricc}}=2$. Several coordinate times during the scalarization are plotted, capturing the evolution just before the code breaks down due to a loss of hyperbolicity. Note that the time frames are not equally spaced to better demonstrate the development of a negative $G_{\rm eff}$ region. In each figure, both $x-y$ and $x-z$ slices are depicted. The apparent horizon is plotted as a white dashed line. \textit{(top)} Time evolution of the scalar field. \textit{(bottom)}  Time evolution of the normalized determinant of the effective metric $G_{\rm eff}$ defined by eq. \eqref{eq:Geff}. Negative values of $G_{\rm eff}$ are depicted in black.}
		\label{fig:hyperbolicity_loss_snapshots}
	\end{figure}
 
	Let us also examine how the evolution of a single black hole looks in the event of a hyperbolicity loss. Several snapshots of the time evolution are depicted in Fig. \ref{fig:hyperbolicity_loss_snapshots}, where the top panel represents the scalar field development while in the lower panel one can see the normalized determinant of the effective metric $G_{\rm eff}$ defined by eq. \eqref{eq:Geff}. The snapshots are adjusted in such a way that the first one is when the scalar field starts developing and is already non-negligible while the last snapshot is the time step just before the code ``crashes''.

	In the plots of the determinant $G_{\rm eff}$, negative values are depicted with black color. Let us remind the reader that inside the black hole horizon (the dashed white line) we have turned off the Gauss-Bonnet coupling, practically setting $\lambda(\varphi)=0$ and $\beta(\varphi)=0$ in the vicinity of the singularity. For the particular models in Fig. \ref{fig:hyperbolicity_loss_snapshots} the cutoff is set at a coordinate radius of roughly $r/M\cong 0.9$ and, after that, the Gauss-Bonnet term is slowly turned on before the horizon is reached (at $r/M\cong 1.07$). This ensures that the Gauss-Bonnet term is turned on completely inside the horizon and that this transition region is far enough from the apparent horizon because, otherwise, some undesired numerical error might propagate outside it \cite{AresteSalo:2022hua,AresteSalo:2023mmd}.  Therefore, only the spacetime outside the apparent horizon is a self-consistent solution of the full field equations and $G_{\rm eff}\cong 1$ deep inside the black hole (i.e. the determinant of the effective metric \eqref{eq:Geff_NonNormalized} is the same as in GR). 
 
    The most important fact that we observe in the graph is the development of a $G_{\rm eff}<0$ (black) region just before the evolution stops. This is a very strong argument that the breakdown of the code is caused by a hyperbolicity loss in the gravitational sector of physical modes (governed by the effective metric \eqref{eq:Geff_NonNormalized}). Therefore, similarly to pure sGB gravity, it is unlikely that this can be improved by a gauge transformation.

    We point out that the loss of hyperbolicity in Fig. \ref{fig:hyperbolicity_loss_snapshots} clearly happens inside the apparent horizon. Actually, what typically happens is that a non-hyperbolic region forms inside the black hole horizon, it grows and expands outside it \cite{Corelli:2022pio}, rendering the solution non-hyperbolic\footnote{Note that if we have a non-hyperbolic region inside the horizon this does non necessarily mean that the solution is non-physical since this problematic region is casually disconnected from the rest of the spacetime.}. We cannot follow such growth, though, because the code crashes right after the determinant of the effective metric $g^{\mu\nu}_{\rm eff}$ turns negative anywhere in the computational domain. Therefore, what is actually observed in simulations, including Fig. \ref{fig:hyperbolicity_loss_snapshots}, is that hyperbolicity loss appears right above the region where we turn on the Gauss-Bonnet term even if this is below the apparent horizon. We have checked that, when moving the cutoff radius further inside or outside, hyperbolicity loss still happens for slightly shifted threshold values of the parameters. Nevertheless, the main qualitative features reported here remain unchanged.

	\subsection{Rotating black holes with Ricci scalar coupling}
	\begin{figure}
		\centering
		\includegraphics[width=0.6\linewidth]{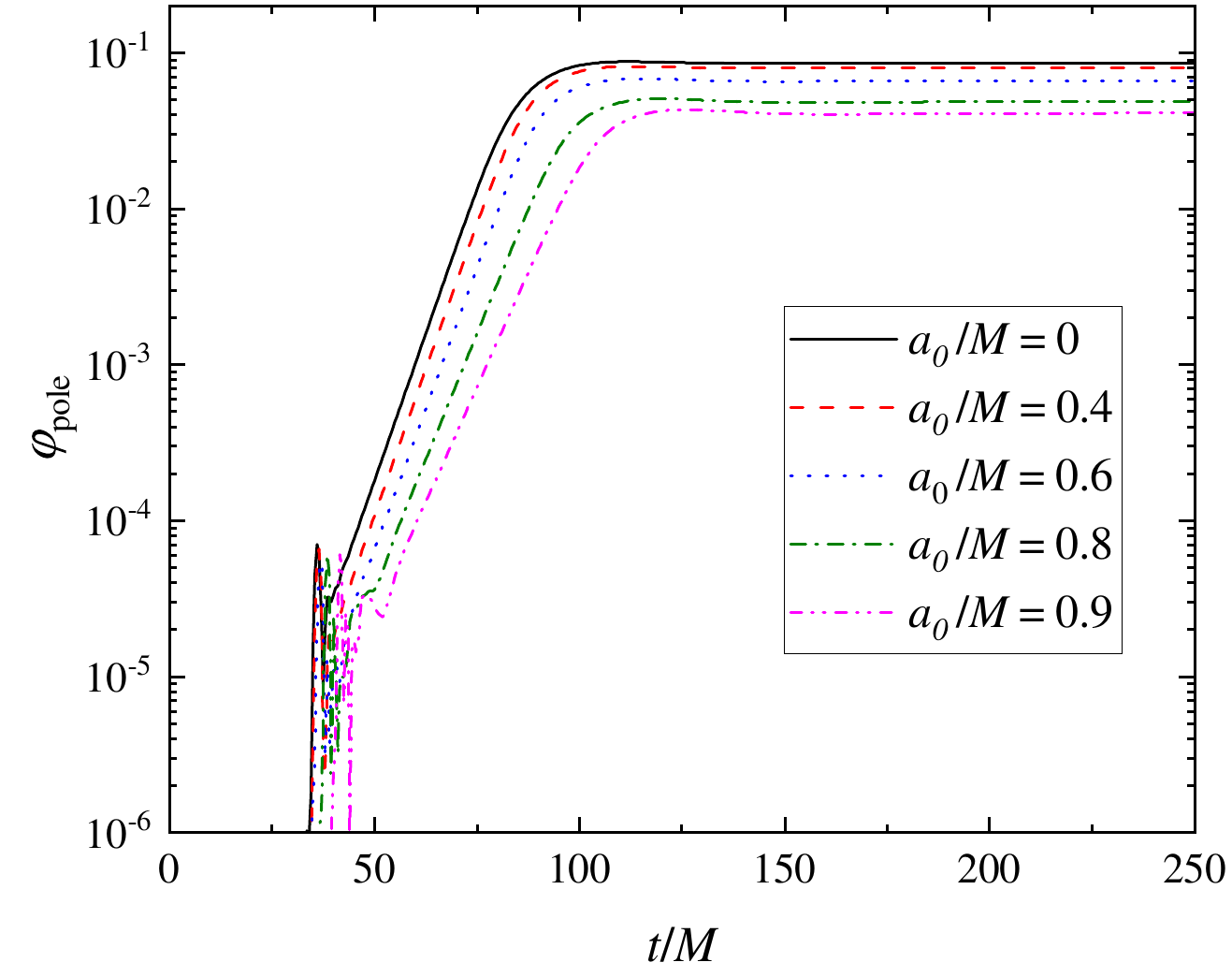}
		\caption{The scalar field on the pole for models with increasing angular momentum $a_0/M$, having $M/\sqrt{\lambda_{\text{GB}}}=0.45$, $\gamma_{\text{GB}}=0$ and $\beta_{\text{Ricc}}=8$.}
		\label{fig:evol_increasing_a}
	\end{figure}
 
 A natural question to ask is whether hyperbolicity is preserved for the models depicted in Fig. \ref{fig:hyperbolic_seq} in case one includes rotation. 
	For that purpose, we have chosen a model from Fig. \ref{fig:hyperbolic_seq} not far away from the point of hyperbolicity loss and performed evolutions for gradually increasing black hole angular momentum. 
	The time evolution of the scalar field on the pole is depicted in Fig. \ref{fig:evol_increasing_a}, which shows that we can perform stable evolution even for very rapidly rotating black holes.
	The maximum depicted value of $a_0/M$ is $0.9$. 
	Above that, i.e. close to the extremal limit, we could also perform evolution of the black hole scalarization but in these cases ending up with a stable hairy black hole requires a subtle adjustment of the auxiliary simulation parameters \cite{AresteSalo:2022hua,AresteSalo:2023mmd}.

	It is interesting to note that the domain of existence of scalarized rotating black holes in sGB gravity presented so far in the literature \cite{Cunha:2019dwb,Collodel:2019kkx} (excluding the case of spin-induced scalarization \cite{Dima:2020yac,Doneva:2020nbb,Herdeiro:2020wei,Berti:2020kgk,Fernandes:2024ztk}) seem to be vanishingly small at a moderate $a_0/M$ due to the violation of the regularity condition. Our simulations suggest that, at least for the considered values of the parameters, scalarized black holes with non-negligible scalar field strength might exist up until (or close to) the extremal limit. Of course, we are considering different coupling functions and a Ricci scalar coupling compared to \cite{Cunha:2019dwb}. In addition, the time evolution we perform can not be a rigorous proof of the existence of stationary black hole solutions. Still, our results suggest that rotating scalarized black holes, where the scalarization is driven by the spacetime curvature rather than the spin of the black hole, exist up until close to the extremal limit. It is also highly likely that this is not attributed to the Ricci coupling alone, but perhaps a good choice of the coupling function in sGB gravity can lead to the same behavior.

	\subsection{Comparison between $\varphi^4$ term and Ricci coupling}
 	\begin{figure}
		\centering
		\includegraphics[width=0.45\linewidth]{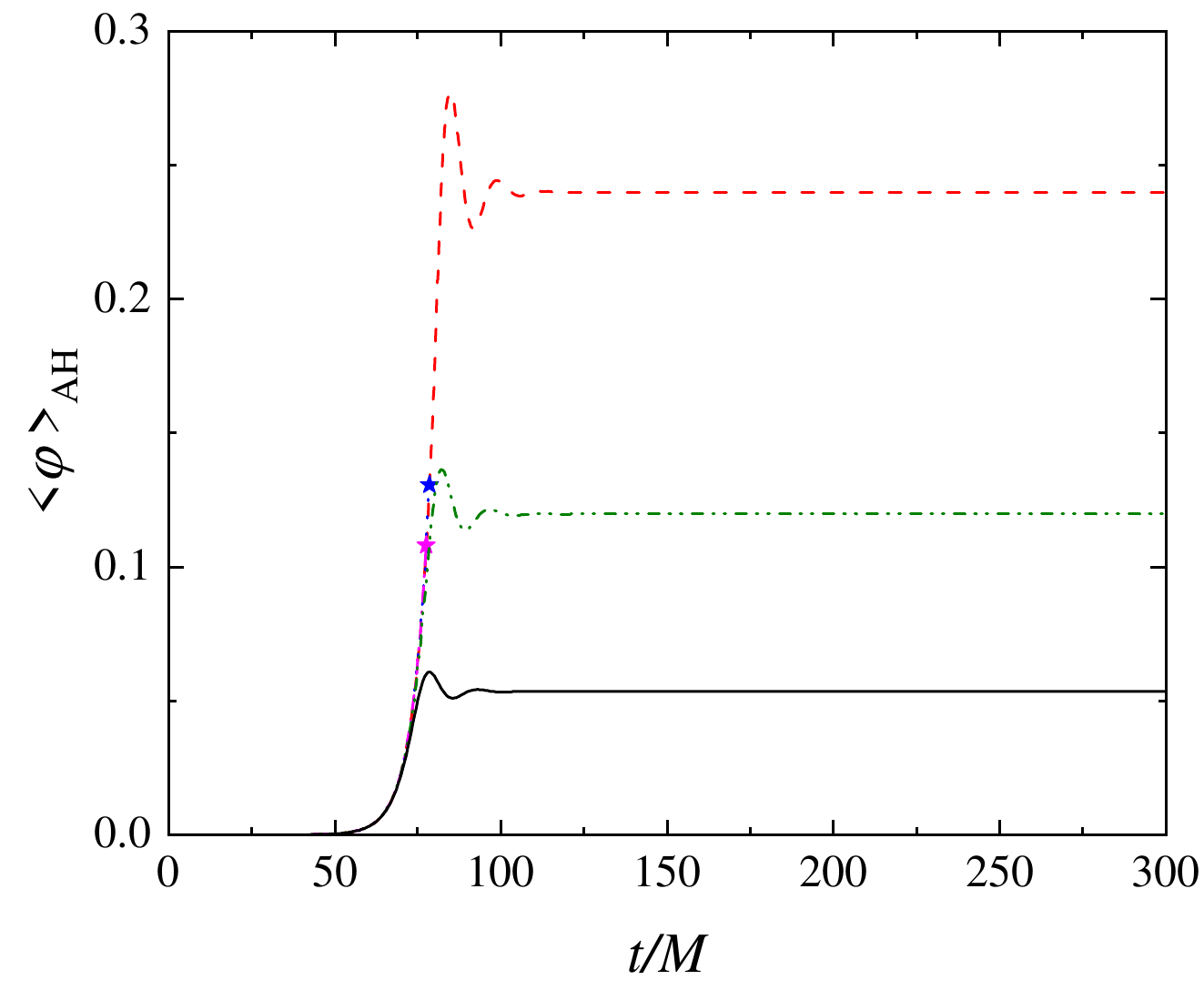}
		\includegraphics[width=0.45\linewidth]{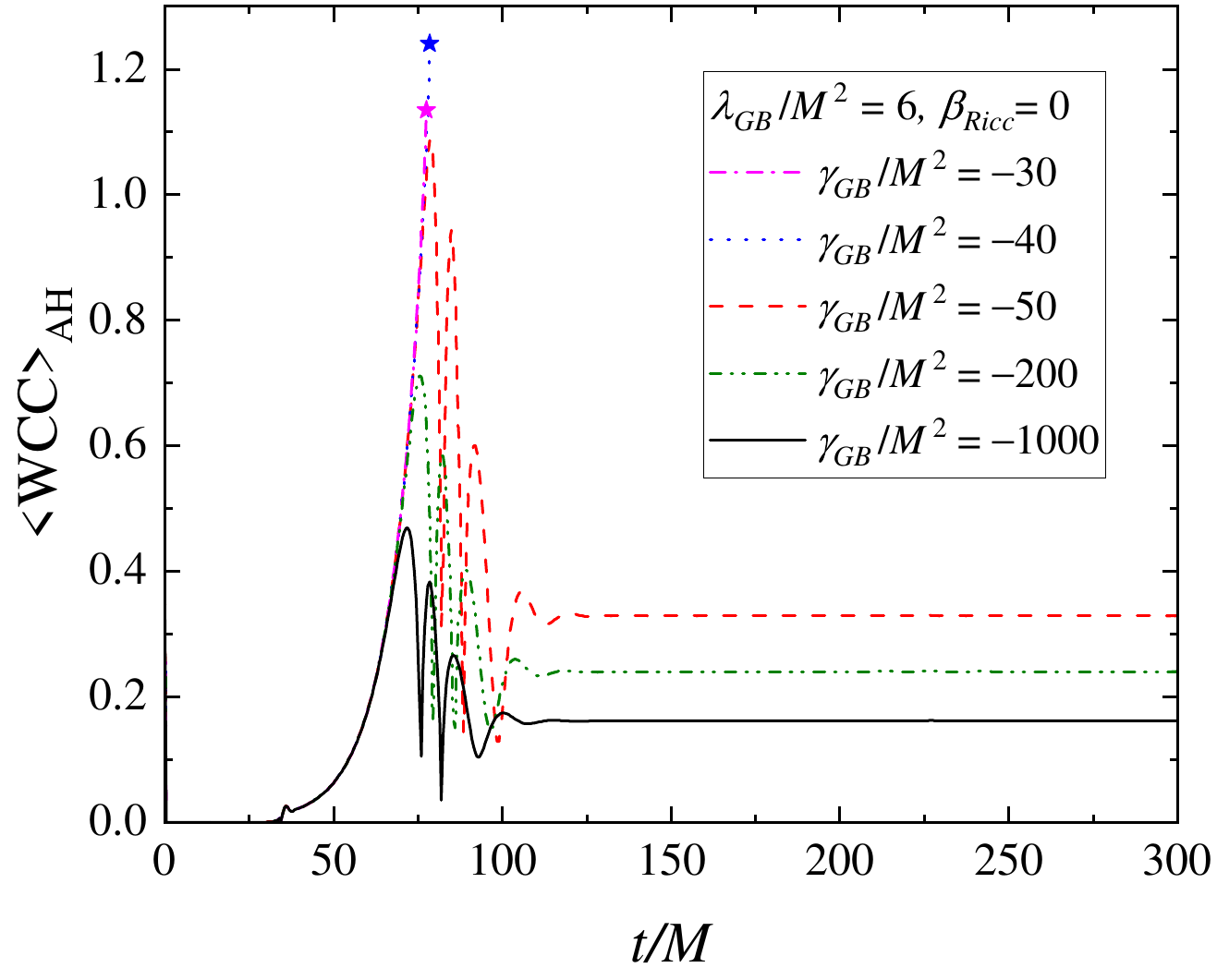}
		\includegraphics[width=0.45\linewidth]{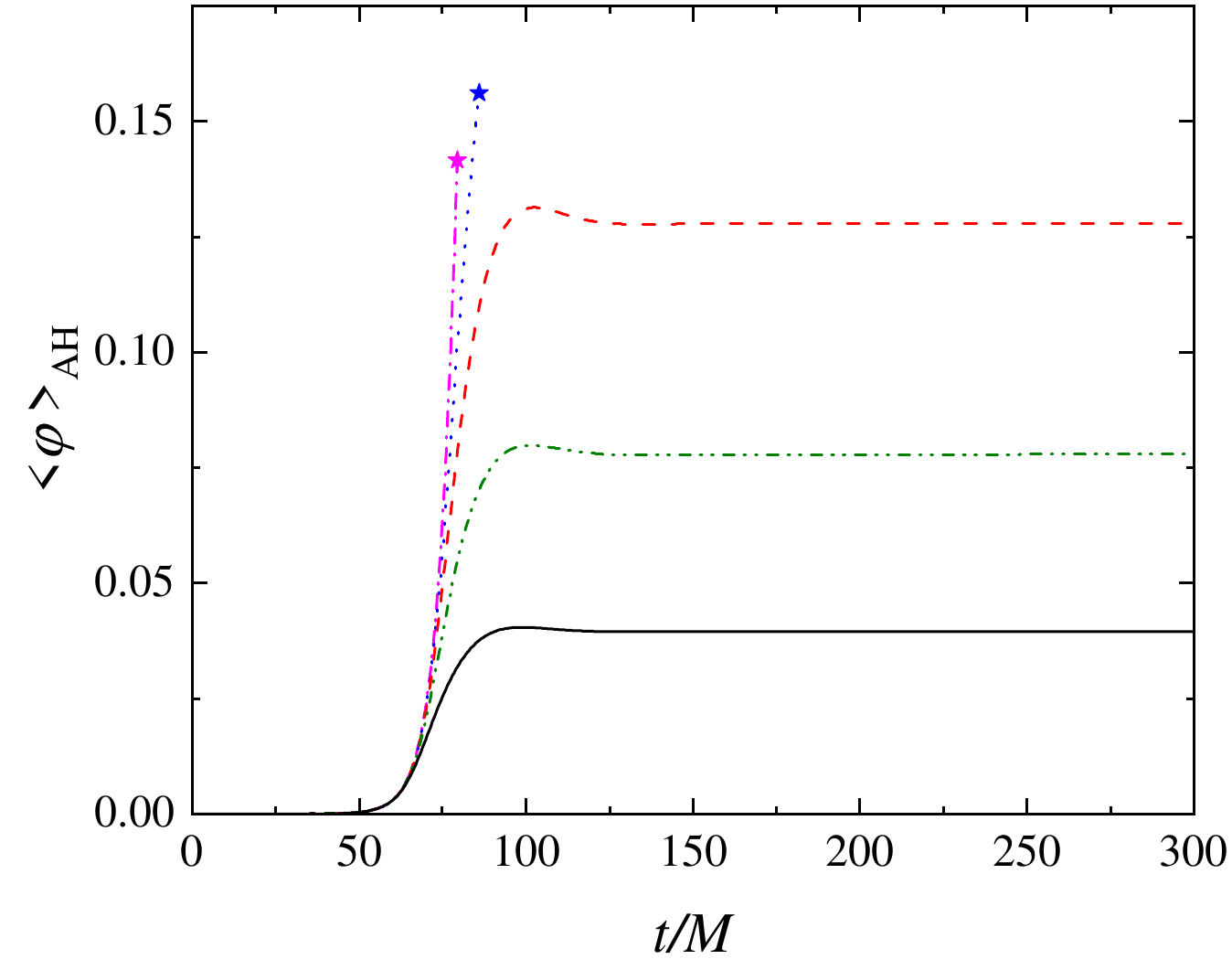}
		\includegraphics[width=0.45\linewidth]{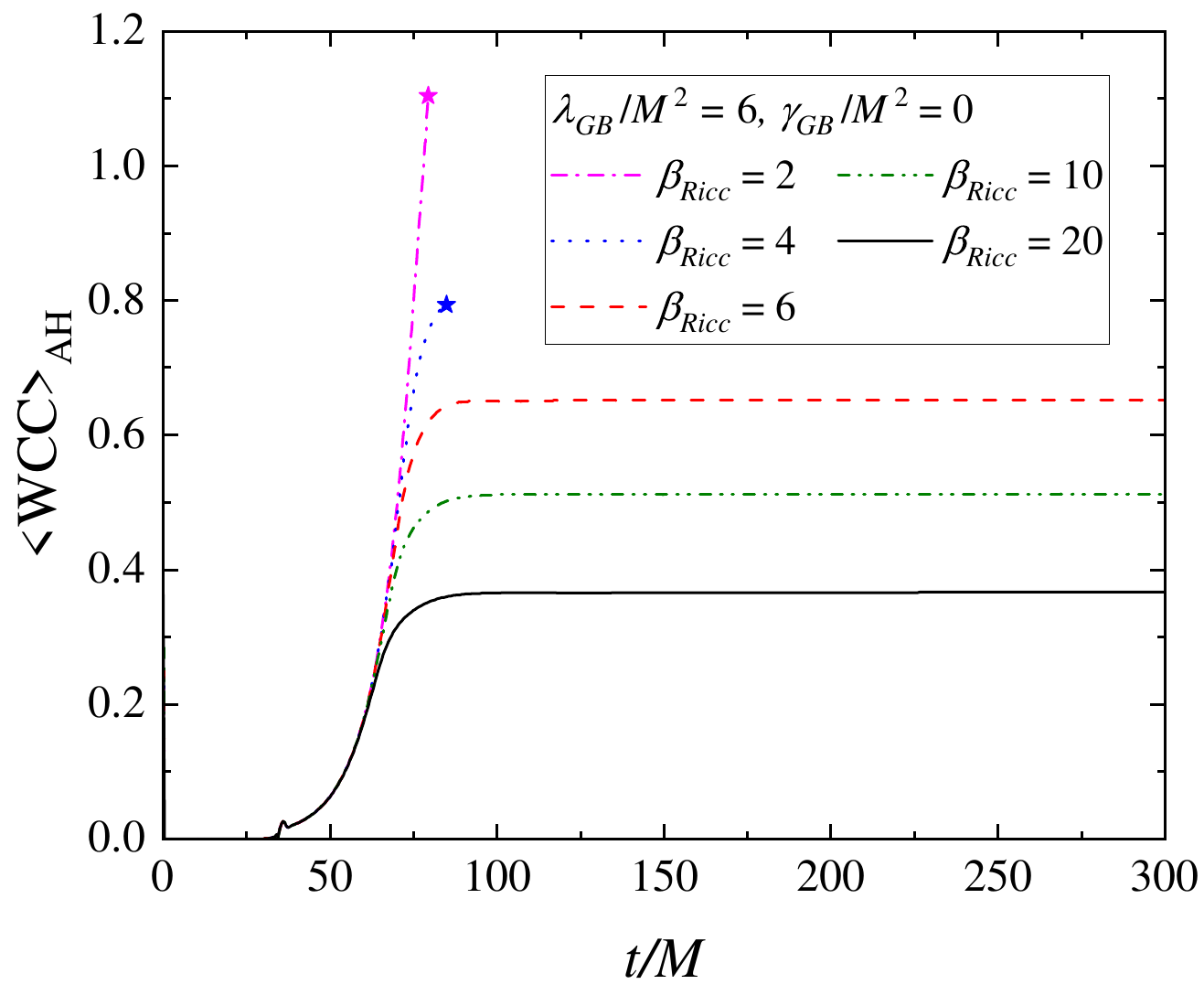}
		\caption{A comparison between the black hole evolution for models in pure scalar-Gauss-Bonnet gravity with the coupling function \eqref{eq:f_phi} (\textit{top panels}) and when including an additional Ricci coupling (\textit{bottom panels}). The \textit{left figures} depict the scalar field evolution for black holes before and after the loss of hyperbolicity. Stars indicate the moment of the evolution when hyperbolicity is lost, which typically happens when the scalar field starts growing during the spontaneous scalarization. The \textit{right figures} demonstrate the evolution of the weak coupling condition defined by eq. \eqref{eq:wcc}. Interestingly, for both theories, the maximum values of the scalar field at the apparent horizon and the maximum of the weak coupling condition, before loss of hyperbolicity is observed, are relatively similar.}
		\label{fig:phi4_Ricc_compare}
	\end{figure}
 
	One of the most important effects of the Ricci coupling term on the spectrum of black hole solutions is that it manages to ``stabilize'' them and with the increase of $\beta_{\text{Ricc}}$ the scalar field gets more and more suppressed. But  this is also exactly the effect that a $\varphi^4$ term has when added to the $\lambda(\varphi)$ coupling function in \eqref{eq:f_phi}. Of course, the two theories are intrinsically different but it will be interesting to compare the loss of hyperbolicity for both of them. We have already pointed out that the loss of hyperbolicity is mainly controlled by the effective metric $g^{\mu\nu}_{\rm eff}$, which differs slightly in the case with and without a Ricci coupling. It might be interesting to ask how far away from the weak coupling condition one can deviate before the modified gauge \cite{Kovacs:2020pns,Kovacs:2020ywu} can no longer secure hyperbolic evolution and how strong the scalar field would be. 
 
 Such a comparison is made in Fig. \ref{fig:phi4_Ricc_compare}, where the time evolution of the scalar field and the weak coupling condition defined by \eqref{eq:wcc} is plotted for models with fixed $M/\sqrt{\lambda_{\text{GB}}}$. The simulations are performed for non-rotating black holes. In the upper panel, $\beta_{\text{Ricc}}=0$ and $\gamma_{\text{GB}}$ is varied (thus we are in sGB gravity with a quadratic and quartic coupling) and in the lower panel $\gamma_{\text{GB}}=0$ while $\beta_{\text{Ricc}}$ varies (Ricci-coupled sGB theory). The ranges of $\gamma_{\text{GB}}$ and $\beta_{\text{Ricc}}$ are chosen on the threshold of hyperbolicity loss. A star at the end of some lines marks hyperbolicity loss while for the rest we observe a saturation of the scalar field to a constant. As one can see, in the upper panels the behavior of the weak coupling condition is oscillatory at early times, which is an artifact of the changes in the scalar field gradient before it settles to an equilibrium value.
	
	In both cases, one can go beyond the weak coupling condition while still maintaining hyperbolicity, and the weak coupling condition defined by \eqref{eq:wcc} reaches the order of unity before hyperbolicity is lost. In the case with the Ricci coupling, one is able to reach a larger scalar field before hyperbolicity is lost but the maximum values for the two theories are still of the same order. Of course, this might change from model to model (e.g. when changing $M/\sqrt{\lambda_{\text{GB}}}$). However, basing ourselves on these results, one can conclude that both theories perform similarly in terms of hyperbolicity loss.

\begin{figure}
		\centering
		\includegraphics[width=0.45\linewidth]{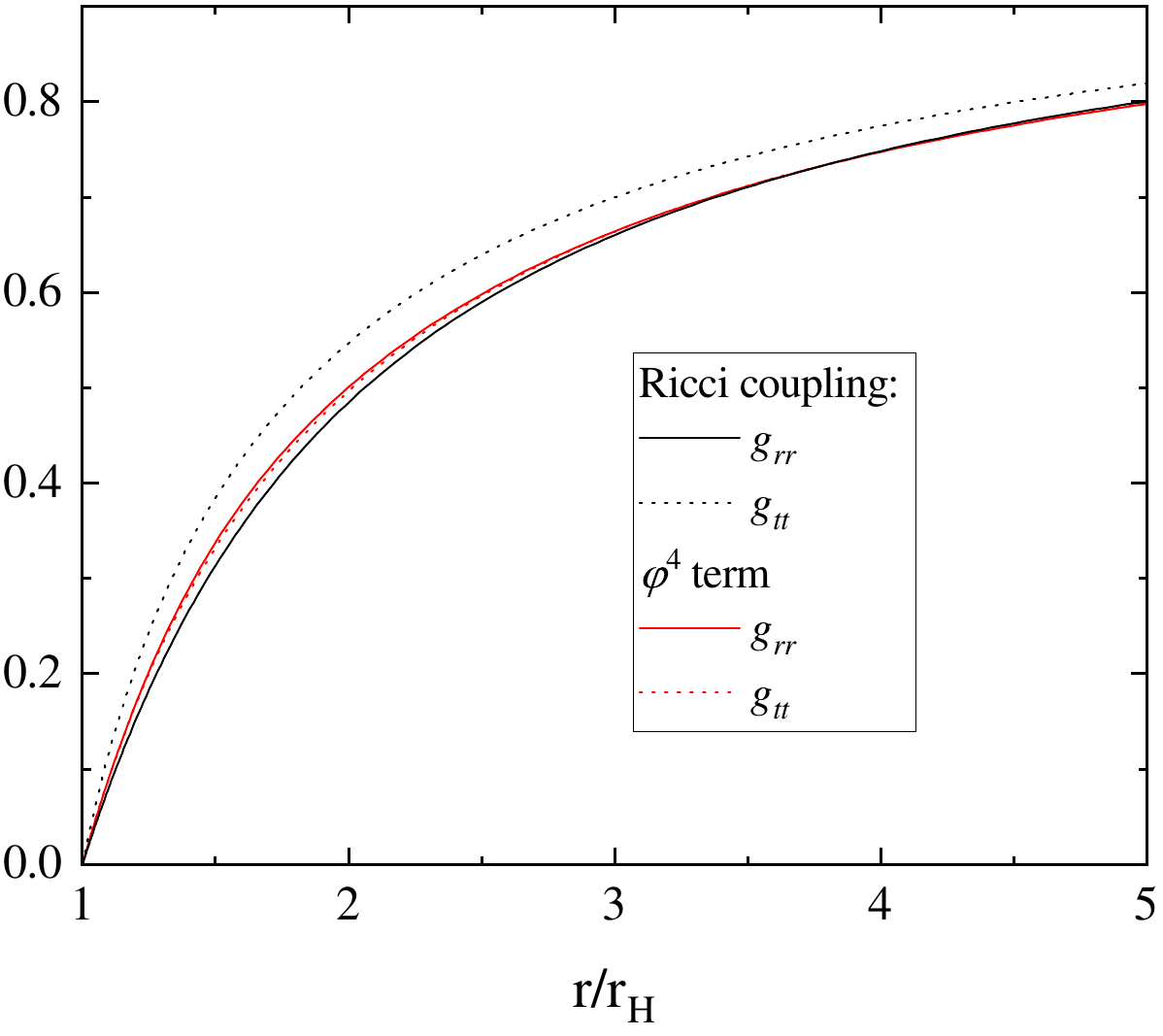}
		\includegraphics[width=0.48\linewidth]{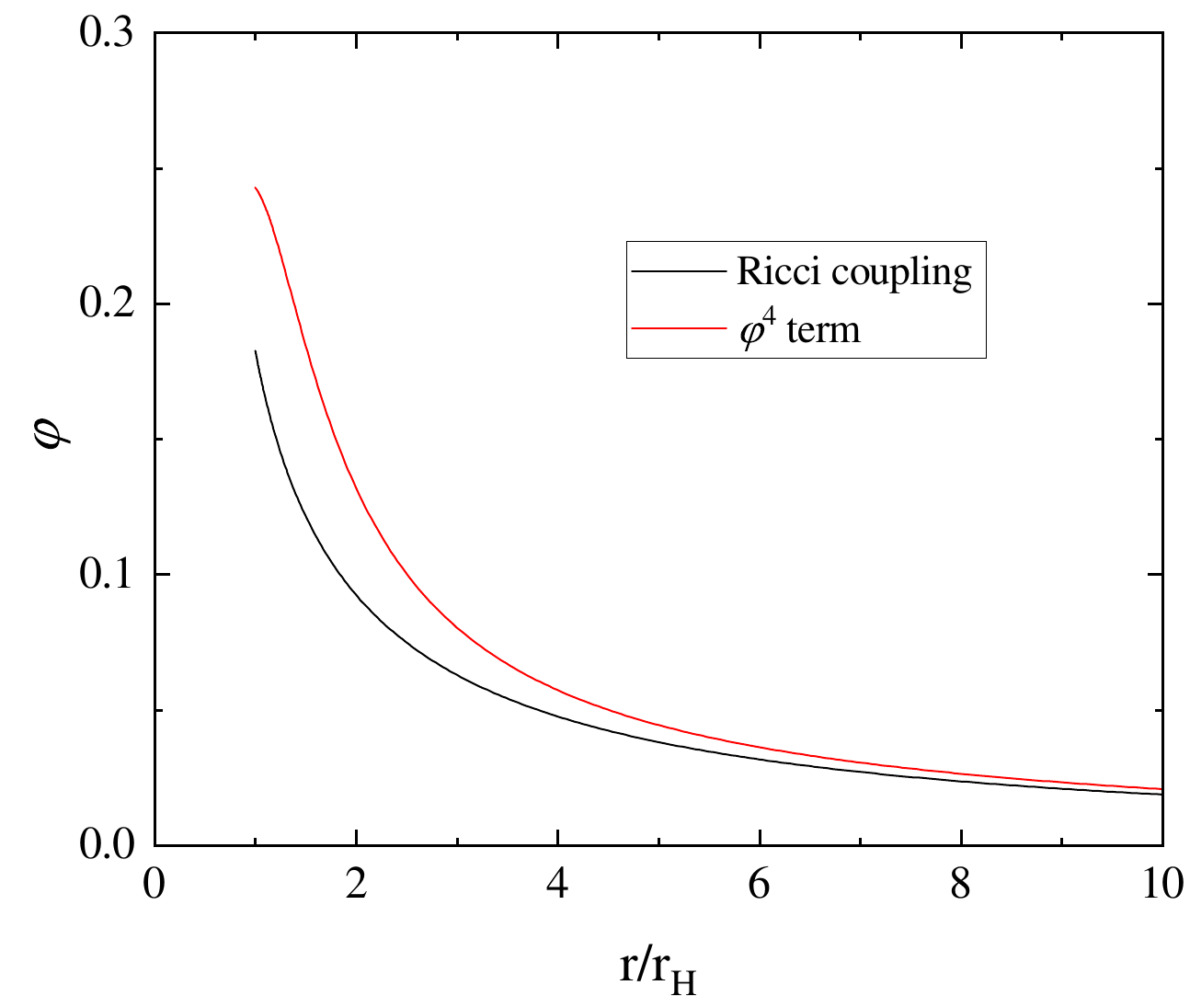}
		\caption{A comparison between the metric and scalar field radial profiles for two static black hole solutions adjusted with the same $M/\sqrt{\lambda_{\text{GB}}}=0.248$ and $D/\sqrt{\lambda_{\text{GB}}}=0.049$, with $D$ being the scalar field charge. One of the solutions is in an sGB theory with Ricci coupling where $\beta_{\text{Ricc}}=5$ and $\gamma_{\text{GB}}=0$, while the other one considers a $\varphi^4$ GB coupling with $\gamma_{\text{GB}}=50$ and $\beta_{\text{Ricc}}=0$.}
		\label{fig:profiles}
	\end{figure}

	Because of these similarities, both in the evolution and the behavior of the spectrum of solutions, one can ask whether the two theories lead to black holes that can be distinguished through observations. For that purpose, we examined the radial profiles of the metric and the scalar field for two models in sGB gravity with and without Ricci scalar coupling in Fig. \ref{fig:profiles}. The parameters of the model are adjusted in such a way that the masses and the scalar charge in the two theories are identical. As one can see, away from the horizon the two solutions look very similar but the differences close to the horizon can be significant. Of course, in this figure we examine only static solutions and it is yet unknown whether the non-linear dynamical will differ as well. Such a study is underway.

    \section{Conclusions}
    In the present paper, we have examined the $3+1$ non-linear evolution of static and rotating black holes in scalar-Gauss-Bonnet gravity with an additional coupling between the scalar field and the Ricci scalar. The study was motivated by the recently discovered nice properties of this theory, such as having general relativity as a late-time cosmology attractor and being able to stabilize hairy black hole solutions that are otherwise unstable in certain flavors of pure sGB gravity \cite{Antoniou:2021zoy}. Extending previous results on hyperbolic spherically symmetric scalar field collapse in sGB gravity with a Ricci coupling \cite{Thaalba:2023fmq}, we explored in detail the well-posedness of the equations of motion in $3+1$ evolutions. For that purpose, a modification of the \texttt{GRFolres} code (based on \texttt{GRChombo}) was developed in order to handle a self-consistent coupled evolution of the field equations.

    The results show that, as expected from the mathematical analysis, the modified gauge developed in sGB gravity \cite{Kovacs:2020pns,Kovacs:2020ywu,AresteSalo:2022hua} also leads to a hyperbolic evolution when adding a Ricci coupling as long as the weak coupling condition is satisfied. As a matter of fact, well-posedness is numerically preserved even slightly above the threshold corresponding to violation of the weak coupling condition. This applies to both static and rotating black holes. As a byproduct of our studies, we have discovered that rotating black holes with a scalar field sourced by the curvature of the spacetime exist for very large angular momenta, close to the extremal limit. This is in contrast with previous studies \cite{Cunha:2019dwb} in sGB gravity, where the domain of existence of black holes was getting really narrow as the extremal limit was approached due to a violation of the regularity condition at the horizon. Our results suggest that with a proper choice of the coupling function between the scalar field and the Gauss-Bonnet invariant, similar near-extremal scalarized black holes with a non-negligible scalar field also exist in pure sGB gravity. A systematic study of the stationary solutions in this case is underway.

    Finally, we have compared the results for the threshold of hyperbolicity loss in the Ricci-coupled sGB theory and also in sGB gravity with a more sophisticated coupling function, possessing both quadratic and quartic scalar field terms. Such a coupling has also a stabilization effect on the scalarized solution even for zero Ricci coupling. Our findings confirm that while the two theories are quite different, the threshold for hyperbolicity loss, in terms of scalar field strength and violation of the weak coupling condition, are very similar. In addition, the profiles of the spacetime metric and the scalar field in the black hole solutions are alike in both cases, even though important differences can be present in the near vicinity of the horizon. It is therefore interesting to study to what extent future observations, e.g. of gravitational waves emitted by merging black holes, will be able to distinguish between both.
         
    \section*{Acknowledgements}
    This study is in part financed by the European Union-NextGenerationEU, through the National Recovery and Resilience Plan of the Republic of Bulgaria, project No. BG-RRP-2.004-0008-C01. DD acknowledges financial support via an Emmy Noether Research Group funded by the German Research Foundation (DFG) under grant no. DO 1771/1-1. LAS is supported by an LMS Early Career Fellowship. We thank Nicola Franchini and Farid Thaalba for useful comments on the draft. We also thank Miguel Bezares and Thomas Sotiriou for useful discussions. We acknowledge Discoverer PetaSC and EuroHPC JU for awarding this project access to Discoverer supercomputer resources. We thank the entire \texttt{GRChombo} \footnote{\texttt{www.grchombo.org}} collaboration for their support and code development work.
    
	\appendix
	\section{Equations of motion in $3+1$ form}\label{app:eom}
	The $3+1$ form of the Einstein equations in our formalism with an arbitrary $T^{\mu\nu}$ yield
	\begin{subequations}\label{eqsccz4}
		\begin{eqnarray}
			\partial_{\perp}\tilde{\gamma}_{ij} &=& -2\alpha\tilde{A}_{ij}+2\tilde\gamma_{k(i}\partial_{j)}\beta^k-\tfrac{2}{3}\tilde{\gamma}_{ij}\partial_k\beta^k, \\
			\partial_{\perp}\chi &=& \tfrac{2}{3}\chi\big(\alpha K - \partial_k\beta^k\big), \\
			\partial_{\perp}K&=&-D^iD_i\alpha +\alpha\left[R+2\,D_iZ^i +K(K-2\Theta)\right] -3\,\kappa_1(1+\kappa_2)\,\alpha\,\Theta+\tfrac{\kappa\,\alpha}{2}\big[S-3\,\rho\big]\nonumber\\
			&&\textstyle-\frac{3\,\alpha\,b(x)}{4(1+b(x))}\Big[R-\tilde{A}_{ij}\tilde{A}^{ij}+\frac{2}{3}K^2 -2\kappa_1(2+\kappa_2)\,\Theta -2\,\kappa\,\rho\Big]\,,\\
			\partial_{\perp}\Theta &=&{\textstyle\frac{\alpha}{2}}\big[R-\tilde A_{ij}\,\tilde A^{ij}+{\textstyle\frac{2}{3}}\,K^2 
			+2\,D^iZ_i-2\,\Theta\,K\big]-Z_i\,D^i\alpha-\kappa_1\big(2+\kappa_2\big)\,\alpha\,\Theta-\kappa\,\alpha\,\rho \nonumber\\
			&&-\textstyle\frac{b(x)}{1+b(x)}\Big\{ \tfrac{\alpha}{2}\big( R-\tilde A_{ij}\,\tilde A^{ij}+{\textstyle\frac{2}{3}}\,K^2 \big)  -\kappa_1\big[2+\kappa_2\big]\,\alpha\,\Theta - \kappa\,\alpha\,\rho
			\Big\}\,,\\
			\partial_{\perp}\tilde{A}_{ij}&=&~\alpha[\tilde{A}_{ij}(K-2\Theta)-2\,\tilde{A}_{ik}\tilde{A}^k_{~j}]+2\tilde{A}_{k(i}\partial_{j)}\beta^k - \tfrac{2}{3}(\partial_k\beta^k)\tilde{A}_{ij} \nonumber\\&&+\chi\left[\alpha\left(R_{ij} + 2D_{(i}Z_{j)}-\kappa\,S_{ij}\right)-D_iD_j\alpha\right]^{\text{TF}} \,,\\
			\partial_{\perp}\hat\Gamma^i
			&=&2\,\alpha\big(\tilde\Gamma^i_{\phantom i kl}\tilde A^{kl}-{\textstyle\frac{2}{3}}\tilde\gamma^{ij}\partial_jK-{\textstyle\frac{3}{2\,\chi}}\,\tilde A^{ij}\partial_j\chi\big) -2\,\tilde A^{ij}\partial_j\alpha-\hat\Gamma^j\partial_j\beta^i + {\textstyle\frac{2}{3}}\,\hat\Gamma^i\partial_j\beta^j + \tfrac{1}{3}\,\tilde\gamma^{ik}\partial_k\partial_j\beta^j + \tilde\gamma^{jk}\partial_j\partial_k\beta^i \nonumber\\
			&&+2\,\alpha\,\tilde\gamma^{ij}\big(\partial_j\Theta- {\textstyle\frac{1}{\alpha}}\,\Theta\,\partial_j\alpha - \tfrac{2}{3}\,K\,Z_j\big) -2\,\kappa_1\,\alpha\,\tilde\gamma^{ij}Z_j\,-2\,\kappa\,\alpha\,\tilde\gamma^{ij}J_j \nonumber\\&&-\textstyle\frac{2\alpha\,b(x)}{1+b(x)}\Big[
			\tilde D_j\tilde A^{ij}-\big(\tfrac{2}{3}\big)\tilde\gamma^{ij}\partial_jK-\tfrac{3}{2\,\chi}\tilde A^{ij}\partial_j\chi %
			+\tilde\gamma^{ij}\big(\partial_j\Theta-\tfrac{1}{3}\,K\,Z_j\big) \nonumber\\&&\hspace{2cm}- \tilde A^{ij}Z_j -\kappa_1\,\tilde\gamma^{ij}\,Z_j-\kappa\,\tilde\gamma^{ij}J_j
			\Big]\,,
		\end{eqnarray}
	\end{subequations}
	where $\partial_{\perp}=\partial_t-\beta^i\partial_i$. Taking into account a scalar field with no potential, with an arbitrary $\beta(\varphi)$  coupling to the Ricci scalar and a non-zero contribution of $\lambda(\varphi)=\lambda_{\text{GB}}f(\varphi)$, with an arbitrary coupling constant $\lambda_{\text{GB}}$ and function $f(\varphi)$, the equations become those above with the following decomposition of $T_{\mu\nu}$, 
	\begin{subequations}
		\begin{eqnarray}
			\kappa\,\rho&=&\tfrac{1}{1-\beta(\varphi)}\big(\tfrac{1}{2}\rho^{\varphi}-B+\lambda_{\text{GB}}\,\rho^{\text{GB}}\big)\,,\\ \kappa\,J_i&=&\tfrac{1}{1-\beta(\varphi)}\big(\tfrac{1}{2}J_i^{\varphi}-B_i+\lambda_{\text{GB}}\,J_i^{\text{GB}}\big)\,,\\
			\kappa\,S_{ij}&=&\tfrac{1}{1-\beta(\varphi)}\big(\tfrac{1}{2}S_{ij}^{\varphi}-B_{ij}+\gamma_{ij}(B-B_{nn})+\lambda_{\text{GB}}\,S_{ij}^{\text{GB}}\big)\,,
		\end{eqnarray}
	\end{subequations}
	where the contribution from the kinetic term is given by
	\begin{subequations}
		\begin{eqnarray}
			\hspace{-0.5cm}\rho^{\varphi}&=&\tfrac{1}{2}\big(K_{\varphi}^2+(\partial\varphi)^2\big)\,,\\
			\hspace{-0.5cm} J_i^{\varphi}&=&K_{\varphi}\,\partial_i\varphi\,,\\
			\hspace{-0.5cm} S_{ij}^{\varphi}&=&(\partial_i\varphi)(\partial_j\varphi)+\tfrac{1}{2}\,\gamma_{ij}\big(K_{\varphi}^2-(\partial\varphi)^2\big)\,,
		\end{eqnarray}
	\end{subequations}
	with $(\partial\varphi)^2=\gamma^{ij}(\partial_i\varphi)(\partial_j\varphi)$ and $K_{\varphi}=-\tfrac{1}{\alpha}\partial_{\perp}\varphi$. The elements $B_{nn}$, $B_{ij}$ and $B_i$ coming from the decomposition of $B_{\mu\nu}=\nabla_{\mu}\nabla_{\nu}\beta(\varphi)=\beta'\nabla_{\mu}\nabla_{\nu}\varphi+\beta''\nabla_{\mu}\varphi\nabla_{\nu}\varphi$ yield
	\begin{subequations}
		\begin{eqnarray}
			\hspace{-0.75cm}B_{nn}=n^{\mu}n^{\nu}B_{\mu\nu}&=&\beta''K_{\varphi}^2-\tfrac{\beta'}{\alpha}(D^k\alpha D_k\varphi+\partial_{\perp}K_{\varphi})\,,\\
			\hspace{-0.75cm}B_i=-\gamma^{\mu}_{~i}n^{\nu}B_{\mu\nu}&=&\beta''K_{\varphi}D_i\varphi+\beta'(D_iK_{\varphi}-K^j_{~i}D_j\varphi)\,,\\
			\hspace{-0.75cm}B_{ij}=\gamma^{\mu}_{~i}\gamma^{\nu}_{~j}B_{\mu\nu}&=& \beta''D_i\varphi D_j\varphi+\beta'(D_iD_j\varphi-K_{\varphi}K_{ij})\,,
		\end{eqnarray}
	\end{subequations}
	with $B=\gamma^{ij}B_{ij}$. With regards to the Gauss-Bonnet sector, we define
	\begin{subequations}\label{edgbcomp}
		\begin{eqnarray}
			\rho^{\text{GB}}&=&\tfrac{\Omega M}{2} - M_{kl}\Omega^{kl}\,, \\
			J^{\text{GB}}_i&=&\tfrac{\Omega_iM}{2}-M_{ij}\Omega^j - 2\big(\Omega^j_ {~[i}N_{j]}-\Omega^{jk}D_{[i}K_{j]k}\big)\,, 
		\end{eqnarray}
	\end{subequations}
	with
	\begin{subequations}
		\begin{eqnarray}\label{MNeq}
			\hspace{-0.5cm}M_{ij}&=&R_{ij}+\tfrac{1}{\chi}\left(\tfrac{2}{9}\tilde{\gamma}_{ij}K^2+\tfrac{1}{3}K\tilde{A}_{ij}-\tilde{A}_{ik}\tilde{A}_j^{~k} \right)\,, \\
			\hspace{-0.5cm}N_i&=&\tilde{D}_j\tilde{A}_i^{~j}-\tfrac{3}{2\chi}\tilde{A}_i^{~j}\partial_j\chi-\tfrac{2}{3}\partial_iK\,, \\
			\hspace{-0.5cm}\Omega_i&=&f'\big(\partial_iK_{\varphi}-\tilde{A}^j_{~i}\partial_j\varphi-\tfrac{K}{3}\partial_i\varphi \big)+f''K_{\varphi}\partial_i\varphi\,,\\
			\hspace{-0.5cm}\Omega_{ij}&=&f'\left(D_iD_j\varphi-K_{\varphi}K_{ij}\right)+f''(\partial_i\varphi) \partial_j\varphi\,,
		\end{eqnarray}
	\end{subequations}
	where $N_i$ is the GR momentum constraint, and $\Omega_i$ and $\Omega_{ij}$ come from the $3+1$ decomposition of ${\mathcal C}_{\mu\nu}=\nabla_{\mu}\nabla_{\nu}f(\varphi)$, where $\lambda(\phi)=\lambda_{\text{GB}}\,f(\varphi)$. In addition, we have 
	\begin{subequations}
		\begin{align}
			M^{\text{TF}}_{ij} &\equiv M_{ij}-\tfrac{1}{3}\gamma_{ij}M\,,\\
			\Omega^{\text{TF}}_{ij} &\equiv \Omega_{ij}-\tfrac{1}{3}\gamma_{ij}\Omega\,,
		\end{align}
	\end{subequations}
	where $M=\gamma^{kl}M_{kl}$ is the GR Hamiltonian constraint and $\Omega=\gamma^{kl}\Omega_{kl}$. Finally, the equations of the two additional degrees of freedom are:
	\begin{subequations}
		\begin{eqnarray}
			\partial_{\perp}\varphi &=& -\alpha\, K_\varphi, \\
			\partial_{\perp}K_{\varphi} &=& \alpha(-D^iD_i\varphi + KK_{\varphi}) - (D^i\varphi) D_i\alpha+\beta'Z^{\beta}-\frac{\lambda_{\text{GB}}}{4}f'(\phi){\mathcal R}^2_{\text{GB}}\,.
		\end{eqnarray}
		\label{eq:scalar_eqs}
	\end{subequations}    
	where
	\begin{equation}
		Z^{\beta}=R+K_{ij}K^{ij}+K^2-\tfrac{2}{\alpha}(\partial_{\perp}K+D_iD^i\alpha)\,.
	\end{equation}
	
	All the definitions above enable us to write down the $3+1$ equations of $\tilde{\gamma}_{ij}$, $\chi$, $\Theta$, $\hat{\Gamma}^i$ and $\phi$ with a r.h.s. not depending on the time derivatives of the variables. The rest of the variables (${\tilde{A}}_{ij}$, $K$ and $K_{\varphi}$) include time derivatives in the r.h.s. and that's why we have to specify them with the following linear system,
	\begin{eqnarray}\label{mat_esgb}
		\begin{pmatrix} X^{kl}_{ij} & Y_{ij} & 0 \\ X^{kl}_K & Y_K & -\tfrac{3\beta'}{2(1-\beta(\varphi))} \\ X^{kl}_{K_{\varphi}} & Y_{K_{\varphi}} & 1 \end{pmatrix}
		\begin{pmatrix} \partial_t \tilde{A}_{kl} \\ \partial_tK \\ \partial_tK_{\varphi} \end{pmatrix}=
		\begin{pmatrix} Z_{ij}^{\tilde{A}} \\ Z^K \\ Z^{K_{\varphi}} \end{pmatrix},
	\end{eqnarray}
	where the elements of the matrix are defined as follows,
	\begin{subequations}
		\begin{eqnarray}
			X_{ij}^{kl}&=& \gamma_i^k\gamma_j^l\big(1-\tfrac{\lambda_{\text{GB}}}{3}\Omega \big)+2\lambda_{\text{GB}}\big(\gamma_{(i}^k\Omega_{j)}^{\text{TF},l} -\tfrac{\gamma_{ij}}{3}\Omega^{\text{TF},kl}-\lambda_{\text{GB}} f'^2M^{\text{TF}}_{ij}M^{\text{TF},kl}\big)\,,\\
			X_K^{kl}&=&\tfrac{\lambda_{\text{GB}}}{2\chi}\big(\Omega^{\text{TF},kl}-\lambda_{\text{GB}}f'^2M\,M^{\text{TF},kl}\big)\,, \\
			X_{K_{\varphi}}^{kl}&=&\tfrac{\lambda_{\text{GB}}}{2\chi}f'M^{\text{TF},kl}\,, \\
			Y_{ij}&=&\tfrac{\lambda_{\text{GB}}}{3}\chi\big(-\Omega^{\text{TF}}_{ij}-\tfrac{6f'\beta'M_{ij}^{\text{TF}}}{1-\beta(\varphi)}+\lambda_{\text{GB}}f'^2M\,M^{\text{TF}}_{ij} \big)\,, \\
			Y_K&=&1-\tfrac{\lambda_{\text{GB}}}{3}\big(\Omega+\tfrac{3\,f'\beta'M}{2(1-\beta(\varphi))}-\tfrac{\lambda_{\text{GB}}}{4}f'^2M^2 \big)\,,\\
			Y_{K_{\varphi}}&=&2\beta'-\tfrac{\lambda_{\text{GB}}}{12}f'M\,,
		\end{eqnarray}
	\end{subequations}
	while the terms of the r.h.s. are
	\begin{subequations}
		\begin{eqnarray}
			Z_{ij}^{\tilde{A}}&=&\chi\big[-D_iD_j\alpha+\alpha\left(R_{ij} + 2D_{(i}Z_{j)} -\kappa\,\bar{S}_{ij}\right) \big]^{\text{TF}}+\beta^k\partial_k\tilde{A}_{ij}+2\,\tilde A_{k(i}\partial_{j)}\beta^k-\tfrac{2}{3}\tilde{A}_{ij}(\partial_k\beta^k)\nonumber\\
			&&+\alpha\left[\tilde{A}_{ij}(K-2\Theta)-2\tilde{A}_{il}\tilde{A}^l_{~j}\right]\,,\\
			Z^K&=&\beta^i\partial_iK-D^iD_i\alpha +\alpha\left[R+2\,D_iZ^i +K(K-2\Theta)\right] -3\,\kappa_1(1+\kappa_2)\,\alpha\,\Theta+\tfrac{\kappa\,\alpha}{2}(\bar{S}-3\rho)\nonumber\\
			&&\textstyle-\frac{3\,\alpha\,b(x)}{4(1+b(x))}\Big[R-\tilde{A}_{ij}\tilde{A}^{ij}+\frac{2}{3}K^2 -2\kappa_1(2+\kappa_2)\,\Theta -2\,\kappa\,\rho\Big]\,,\\
			Z^{K_{\varphi}}&=& \beta^i\partial_i K_\varphi+\alpha\big(-D^iD_i\varphi+K\,K_\varphi\big)-(D^i\varphi)D_i\alpha+\alpha\,\beta'\,\bar{Z}^{\beta}-\tfrac{\lambda_{\text{GB}}}{4}\,\alpha\,f'\,\bar{\mathcal R}^2_{\text{GB}}\,,
		\end{eqnarray}
	\end{subequations}
	where the bar denotes that the terms depending on the time derivatives of ${\tilde{A}}_{ij}$, $K$ and $K_{\varphi}$ of the expressions $Z^{\beta}$, $S_{ij}$, $S$ and ${\mathcal R}^2_{\text{GB}}$ are subtracted, yielding
	\begin{subequations}
		\begin{eqnarray}
			\kappa\,\bar{S}_{ij}&=&\tfrac{1}{1-\beta(\varphi)}\big(\tfrac{1}{2}S_{ij}^{\varphi}-B_{ij}+\gamma_{ij}(B-\bar{B}_{nn})+\lambda_{\text{GB}}\bar{S}_{ij}^{\text{GB}}\big)\,,\\
			\bar{\mathcal R}^2_{\text{GB}}&=&-\tfrac{4}{3}\,M\left[-\tfrac{1}{\alpha}\beta^i\partial_iK + \tfrac{1}{\alpha}D_iD^i\alpha-\tilde{A}_{ij}\tilde{A}^{ij}-\tfrac{K^2}{3} \right]-4\, H\ \nonumber\\
			&&+8\,M^{\text{TF},kl}\left[\tfrac{1}{\alpha}D_kD_l\alpha+\tfrac{1}{\chi}\left(\tilde{A}_{kj}\tilde{A}^j_{~l}-\hat{\Theta}_{kl}\right) \right]\,,\nonumber\\
			\bar{Z}^{\beta}&=&R+K_{ij}K^{ij}+K^2+\tfrac{2}{\alpha}(\beta^i\partial_iK-D_iD^i\alpha)\,,
		\end{eqnarray}
	\end{subequations}
	with $\bar{S}=\gamma^{ij}\bar{S}_{ij}$, $\bar{B}_{nn}=\beta''K_{\varphi}^2-\tfrac{\beta'}{\alpha}(D^k\alpha D_k\varphi-\beta^k\partial_kK_{\varphi})$ and
	\begin{subequations}
		\begin{eqnarray}
			\bar{S}_{ij}^{\text{GB,TF}}&=&-\tfrac{1}{3}\left(\Omega^{TF}_{ij}-\lambda_{\text{GB}}f'^2MM^{\text{TF}}_{ij}\right)\big[-\tfrac{1}{\alpha}\beta^i\partial_iK
			+\tfrac{1}{\alpha}D_iD^i\alpha-\tilde{A}_{kl}\tilde{A}^{kl}-\tfrac{K^2}{3} \big]\nonumber\\
			&&-M_{ij}^{TF}\left[\Omega+f''(K_{\varphi}^2-(\partial\varphi)^2)-\beta'f'Z^{\beta}-\lambda_{\text{GB}}f'^2H \right]\nonumber\\
			&&-\tfrac{1}{3}\,\Omega\left[\tfrac{1}{\alpha}D_iD_j\alpha+\tfrac{1}{\chi}\big(\tilde{A}_{im}\tilde{A}^m_{~j}-\hat{\Theta}_{ij}\big) \right]^{\text{TF}}-\tfrac{2}{3}\,\Omega_{ij}^{\text{TF}}\left(\tfrac{1}{\alpha}D_kD^k\alpha-\tilde{A}_{kl}\tilde{A}^{kl} \right) \nonumber\\
			&&+2\,\Omega_{(i}^{\text{TF},k}\left[\tfrac{1}{\alpha}D_{j)}D_k\alpha+\tfrac{1}{\chi}\big(\tilde{A}_{j)m}\tilde{A}_k^{~m}-\hat{\Theta}_{j)k}\big) \right] +\left[N_{(i}\Omega_{j)}\right]^{\text{TF}}\nonumber\\
			&&-2\,\left(\tfrac{1}{3}\,\gamma_{ij}\,\Omega^{\text{TF},kl}+\lambda_{\text{GB}}f'^2M_{ij}^{\text{TF}}M^{\text{TF},kl} \right)\left[\tfrac{1}{\alpha}D_kD_l\alpha
			+\tfrac{1}{\chi}\big(\tilde{A}_{km}\tilde{A}^m_{~l}-\hat{\Theta}_{kl}\big) \right]\nonumber\\
			&&-2\left(D_kA_{ij}-D_{(i}A_{j)k}\right)\Omega^k-\gamma_ {ij}\,(D^kA_{kl})\,\Omega^l+\Omega_{(i}D^kA_{j)k} \,,\\
			\bar{S}^{\text{GB}}&=&\tfrac{2}{3}\left(\Omega-\tfrac{\lambda_{\text{GB}}}{4}f'^2M^2\right)\left[-\tfrac{1}{\alpha}\beta^i\partial_iK+\tfrac{1}{\alpha}D_iD^i\alpha-\tilde{A}_{ij}\tilde{A}^{ij}-\tfrac{K^2}{3} \right]-2\Omega^iN_i-\Omega^{\text{TF},ij}\,M^{\text{TF}}_{ij}\nonumber\\
			&&+2\, M\left(\tfrac{1}{4}\,f''(K_{\varphi}^2-(\partial\varphi)^2)-\tfrac{\beta'}{4}f'Z^{\beta}-\tfrac{\lambda_{\text{GB}}}{4}f'^2H+\tfrac{1}{3}\,\Omega\right) -\rho^{\text{GB}}
			\nonumber\\
			&&+\big(\Omega^{\text{TF},kl}-\lambda_{\text{GB}}f'^2MM^{\text{TF},kl}\big)\left(\tfrac{1}{\alpha}D_kD_l\alpha+\tfrac{1}{\chi}\tilde{A}_{km}\tilde{A}_{~l}^m -\tfrac{\hat{\Theta}_{kl}}{\chi} \right)\,,
		\end{eqnarray}
	\end{subequations}
	where we have used $\hat{\Theta}_{kl}=\tfrac{1}{\alpha}{\mathcal L}_{\beta}\tilde{A}_{kl}+\tfrac{2}{3}\left(K-\tfrac{1}{\alpha}\partial_i\beta^i \right)\tilde{A}_{kl}$ with ${\mathcal L}_{\beta}\tilde{A}_{ij}=\beta^k\partial_k \tilde A_{ij}+2\tilde A_{k(i}\partial_{j)}\beta^k$ and 
	\begin{equation}
		\begin{aligned}
			H=
			-\tfrac{4}{3}D_iK\big(N^i+\tfrac{D^iK}{3}\big) +2\,D_iA_{jk}\big(D^iA^{jk}-D^jA^{ik} \big)-2\,N_i\,N^i\,.
		\end{aligned}
	\end{equation}

	\section{Code testing}\label{app:convergence}
	In this appendix we present the basic evidence for the validity of the developed extension of \texttt{GRFolres}. The first test we have made is to verify that the late-time evolution of \texttt{GRFolres}, namely when the black hole scalarizes and reaches a quasi-equilibrium state, agrees with the results from the solution of the static field equations. This is already presented in Fig. \ref{fig:hyperbolic_seq}, where one can see a very good agreement between the masses and the scalar charges obtained by the modified \texttt{GRFolres} evolution and the static black hole solutions.

\begin{figure}
	\centering
	\includegraphics[width=0.45\linewidth]{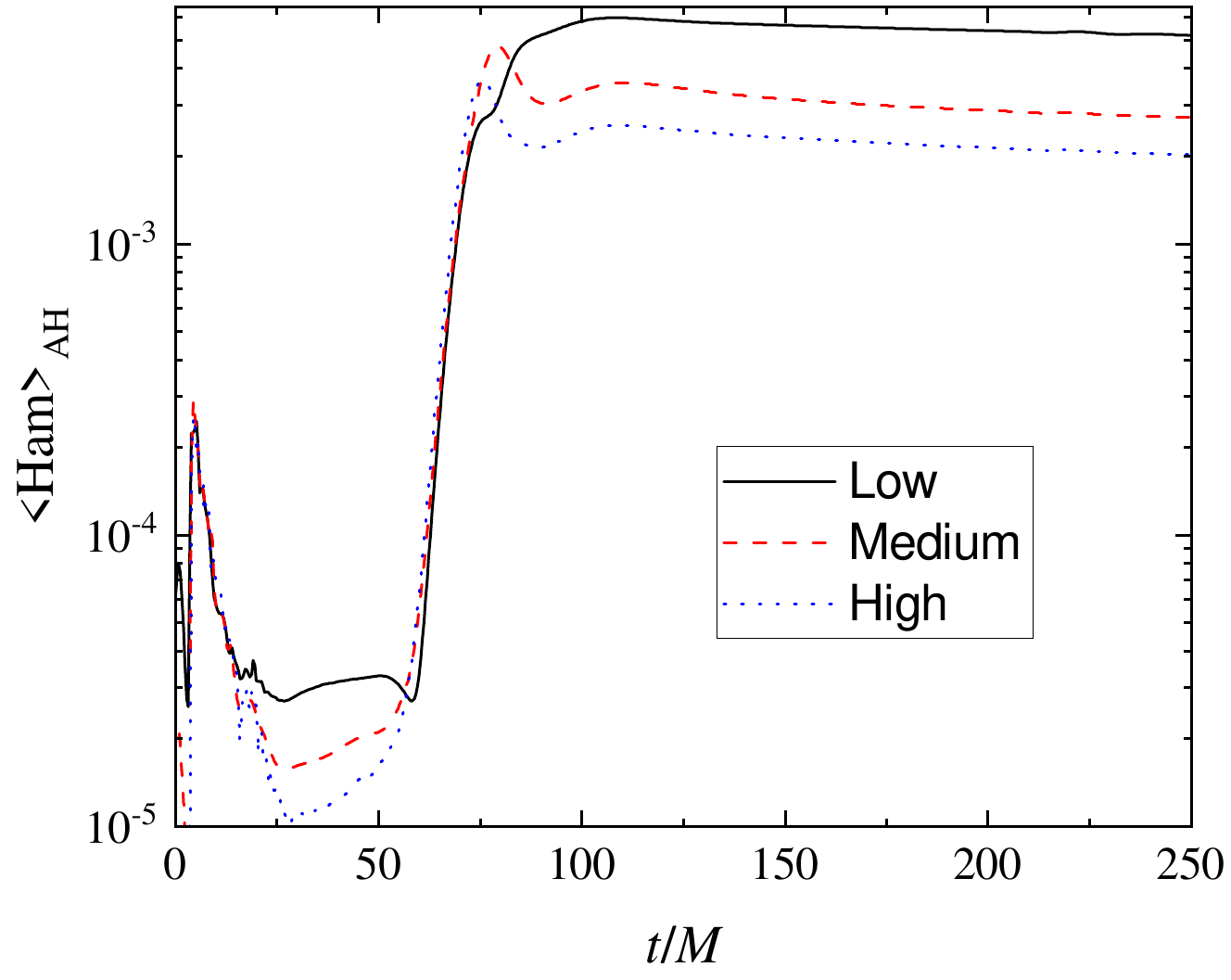}
	\includegraphics[width=0.45\linewidth]{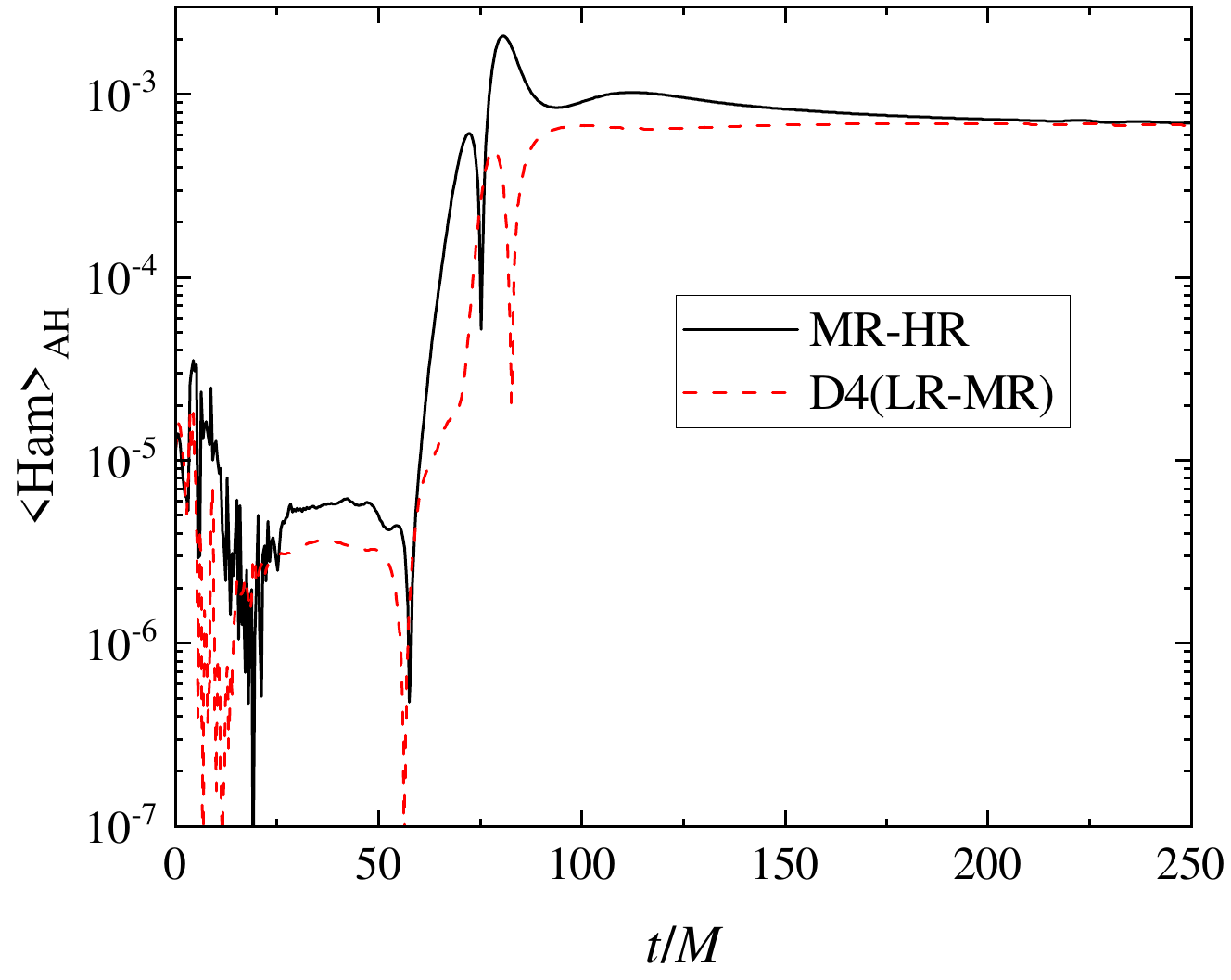}
	\caption{The difference between the scalar field time evolution performed for three different resolutions for $\lambda_{\text{GB}}/M^2=6$, $\gamma_{\text{GB}}=0$, and $\beta_{\text{Ricc}}=10$. The three resolutions are chosen to have 96 (low resolution), 128 (medium resolution), and 160 (high resolution) points at the coarser level in each spatial direction, with 6 refinement levels, and a domain size of $256M$. (\textit{left panel})  The average value of the Hamiltonian constraint at the apparent horizon. (\textit{right panel}) The difference between the average Hamiltonian constraint at the event horizon for medium and high resolution (black solid holes),  low and medium resolution (red dashed line) multiplied by the fourth-order convergence factor $D_4=\frac{h^4_{\rm MR} - h^4_{\rm HR}}{h^4_{\rm LR} - h^4_{\rm ML}}$.}
	\label{fig:Ham_Convergence}
    \end{figure}

    The average value of the Hamiltonian constraint at the apparent horizon, as well as a convergence plot, are presented in Fig. \ref{fig:Ham_Convergence}. We observe that the convergence matches well to a fourth order, which is consistent with the order of the finite difference stencils, as was also shown in the pure sGB case \cite{AresteSalo:2022hua}.
    
	\bibliography{references}

\end{document}